\documentclass[aps,prd,twocolumn,showpacs,superscriptaddress,groupedaddress,nofootinbib]{revtex4}

\usepackage{amsmath}
\usepackage{amssymb}
\usepackage{amsthm}
\usepackage{bm}
\usepackage{dcolumn}
\usepackage{graphicx}
\usepackage{subfigure}

\begin{document}

\title{Exploring tidal effects of coalescing binary neutron stars in numerical relativity}
\author{Kenta Hotokezaka}
\affiliation{Department of Physics,~Kyoto~University,~Kyoto~606-8502,~Japan}

\author{Koutarou Kyutoku}
\affiliation{
Theory Center, Institute of Particles and Nuclear Studies,
KEK, Tsukuba,~305-0801,~Japan}

\author{Masaru Shibata}
\affiliation{Yukawa Institute for Theoretical
Physics,~Kyoto~University,~Kyoto~606-8502,~Japan}

\begin{abstract}
We study gravitational waves emitted in the late inspiral stage of
binary neutron stars by analyzing the waveform obtained in
numerical-relativity simulations.  For deriving the physical
gravitational waveforms from the numerical results, the resolution
extrapolation plays an essential role for our simulations.  The
extrapolated gravitational-wave phases are compared with those
calculated in the post-Newtonian (PN) and effective-one-body (EOB)
formalisms including corrections of tidal effects.  We show that the
extrapolated gravitational-wave phases in numerical relativity agree
well with those by the PN and EOB calculations for most of the
inspiral stage except for a tidally-dominated, final inspiral stage,
in which the PN and EOB results underestimate the tidal
effects. Nevertheless, the accumulated phase difference between our
extrapolated results and the results by the PN/EOB calculations is at most 1--3~radian
in the last 15~cycles. 
\end{abstract}
\maketitle

\section{Introduction}
The inspiral and merger of coalescing compact binaries are among the
most promising sources for kilometer-size laser-interferometric
gravitational-wave detectors~\cite{abbot10, accadia11, hild06}.  A
statistical study based on the stellar evolution synthesis (e.g.,
Refs.~\cite{O'S10,abadie10}) suggests that detection rate $\sim 1 -
100$ $\rm{yr}^{-1}$ may be achieved by advanced detectors such as
advanced LIGO~\cite{aligo}, advanced VIRGO~\cite{avirgo}, and KAGRA
(LCGT)~\cite{kuroda10}, which will be in operation in this decade.

One of the key goals after the first detection of gravitational waves
from inspiraling black hole-neutron star (BH-NS) binaries and binary
neutron stars (NS-NS) achieved in the near future will be to extract
binary parameters such as mass, spin, and radius of each object in the
binary systems.  In particular, the mass and quantities related to the
finite size of the neutron star will provide us the essential
information for the equation of state (EOS) of the neutron-star
matter.  The mass of two neutron stars will be determined with a high
accuracy $\alt 1\%$, if the gravitational-wave signals in the inspiral
stage are detected with the signal-to-noise ratio $\agt
10$~\cite{cutler94}. 

It is more challenging to determine the parameters related to the
finite size of neutron stars.  Several methods to measure such a
quantity by observing the gravitational-wave signals from NS-NS and
BH-NS binaries have been already
proposed~\cite{shibata05,read09b,kiuchi10,sekiguchi11,andersson11,lackey12,bauswein12,bauswein12b}.
In particular, extracting the tidal deformability of neutron stars
from the gravitational-wave signals from NS-NS inspirals is one of the
convincing ways \cite{flanagan08}. 
For realizing this method, one needs to go beyond
the point-particle approximation to model the gravitational waveform
in NS-NS inspirals.  In other words, one has to derive the
gravitational waveform in NS-NS inspirals including tidal effects,
which influence the dynamics of the binary systems in the
late inspiral phase (e.g., Ref.~\cite{mora04}).
When the tidal deformability of a neutron star can be measured by the
gravitational-wave observations, one can constrain on the neutron-star matter EOS.
Once the neutron-star EOS is known, one also can measure the relationship between the luminosity
distance and the redshift of the binary using only the information of the
gravitational waveform through the tidal deformation of the inspiralling neutron stars~\cite{messenger12}
with more refined detectors such as
Einstein telescope \cite{punturo10}.
Therefore, modeling the gravitational waveform in NS-NS inspirals
including tidal effects is important not only from astrophysical
point of view but also from the viewpoint of nuclear physics and
cosmology.

For the early stage of NS-NS inspirals ($f \lesssim 400$~Hz), a post
Newtonian (PN) gravitational waveform was derived by Flanagan and
Hinderer including the leading-order tidal effects~\cite{flanagan08}.
It shows that the tidal interaction affects the evolution of the
gravitational-wave phase only through a single parameter, namely the
tidal deformability of a neutron star, up to the leading order.  They
also showed that the tidal deformability of a neutron star could be
measurable by the advanced gravitational-wave detectors by using the
gravitational-wave signals for 10 -- 400~Hz, if the tidal 
deformability of a neutron star is sufficiently large or if we observe
an event with a high signal-to-noise ratio (see also 
Ref.~\cite{hinderer10}).

More recently, focusing on the late stage of NS-NS inspirals ($f >
400$~Hz), Damour and his collaborators \cite{damour12} explored the
measurability of the tidal deformability with the advanced
gravitational-wave detectors.  They used an effective one body (EOB)
formalism for modeling the waveform in NS-NS inspirals including tidal
effects up to 2PN order. 
They concluded that the tidal deformability of a neutron
star can be measured by the advanced gravitational-wave detectors for
the gravitational-wave signals of which the signal-to-noise ratio is
higher than 16 for any EOS that satisfies the constraint of the maximum
mass $\geq 1.97M_{\odot}$ \cite{demorest10}.  The key assumption of
their study is that the EOB formalism is valid up to the contact point
of the two neutron stars.

In the stage just before the merger, nonlinear hydrodynamics
effects play a crucial role for the evolution of NS-NS
binaries~\cite{lai94}.  In addition, higher-PN tidal
corrections may yield a pole-like behavior of the tidal interactions
near the last unstable orbit~\cite{bini12}.

For better understanding the precise motion and the waveform in this
late inspiral stage, a numerical-relativity (NR) simulation is
probably the best approach (see, e.g., Refs.~\cite{duez10,shibata11,faber12} for
a review of this field).  Recently, long-term simulations for NS-NS
inspirals were performed by three groups~\cite{baiotti11,
  thierfelder11, hotokezaka11} aiming at the derivation of accurate
gravitational waveforms for the late inspiral stage.  Baiotti and
his collaborators performed a NR simulation employing a $\Gamma$-law EOS and
compared the resulting waveforms of the \textit{highest} resolution
simulation with the analytic models calculated in the EOB and Taylor
T4 formalisms~\cite{baiotti10,baiotti11}.  They suggested that the
tidal effects might be significantly amplified by higher-PN
tidal corrections even in the early inspiral phase.

Bernuzzi and his collaborators performed a simulation with
$\Gamma$-law EOS ($\Gamma=2$, and the compactness of a neutron star
is 0.14)~\cite{bernuzzi11,bernuzzi12}.
In Ref.~\cite{bernuzzi11}, they studied the
convergence of the numerical results for NS-NS inspirals.
They concluded that the convergence of the simulation is
second order up to contact. 
They also compared the resulting \textit{extrapolated} waveform with that of the Taylor T4
formalism for the point-particle approximation and for including the tidal corrections.
They found that the accumulated phase difference is about 1.5 radian at
contact for a particular model of the NS-NS binary.
In the subsequent paper~\cite{bernuzzi12},
they compared the waveform derived by the \textit{highest} resolution simulation
with the waveform calculated in the EOB formalism.
They found that the EOB formalism including tidal corrections
up to the next-to-next-to leading order is currently the most robust way to describe
the waveform of NS-NS inspirals.
In addition, they excluded the huge amplification of the tidal corrections
suggested in Ref.~\cite{baiotti11}.

In this paper, we study NS-NS inspirals by NR simulations with
three different EOSs and compare
the \textit{extrapolated} NR waveforms with those calculated in the EOB and
Taylor T4 formalisms.  Here we extrapolate NR data with
a new extrapolation procedure, the time and phase extrapolation.
For studying the dependence of the tidal
effects on the neutron-star matter EOS, we employ a
piecewise-polytropic EOS of Ref.~\cite{read09}, which can
approximately describe the EOS based on nuclear theoretical
calculations and more realistic than $\Gamma$-law EOS adopted in Refs.\cite{baiotti11, bernuzzi11, bernuzzi12}.
In this paper, (i)~we obtain the physical
gravitational-wave phase by extrapolation; (ii)~we then compare the
\textit{extrapolated} waveforms with those of the analytic models calculated in
the Taylor T4 and EOB formalisms; (iii)~we clarify the tidal effects
on the gravitational-wave phase and show the validity of the analytic
modeling in the late inspiral phase. 

The paper is organized as follows.  In Sec.~II, we briefly review the
analytic modeling of gravitational waves emitted from a tidally
interacting binary system.  In Sec.~III, we summarize the formulation
and numerical schemes employed in our numerical code {\tt SACRA}, and
review the EOS employed in this study.  In Sec.~IV, we describe our
method of data analysis of the numerical waveforms; the radius
extrapolation and the resolution extrapolation.  In Sec.~V, we compare
the extrapolated gravitational-wave phase with those derived in the
analytic modeling.  Section~VI is devoted to a summary.  Throughout
this paper, we adopt the geometrical units of $c=G=1$ where
$c$ and $G$ are the speed of light the gravitational
constant respectively.

\section{Tidal effects in a binary system}

In this section, we describe analytic models for the calculation of
gravitational waves emitted from NS-NS inspirals in close orbits.  We
briefly summarize the definition of the tidal deformability of a
neutron star, and the PN and EOB descriptions of the
tidally-interacting dynamics of close NS-NS binaries. 

\subsection{Tidal deformability of a neutron star}

In a close binary system for which the separation between two stars is
a few times larger than the stellar radius, each star is deformed from
its hypothetical equilibrium shape in isolation due to the tidal fields.
Assuming that neutron stars are spherically symmetric in the zeroth order,
such deformation can be described as the linear responses of neutron
stars to external tidal
fields~\cite{hinderer08,damour09,binnington09}, as long as the degree
of the tidal deformation is small.  In this linear theory, one assumes
that the mass quadrupole moment of a star, $Q_{ij}$, is proportional
to the external quadrupolar tidal fields $\mathcal{E}_{ij}$ as,
\begin{eqnarray}
Q_{ij} = -\lambda \mathcal{E}_{ij},
\end{eqnarray}  
where $\lambda$ is the quadrupolar tidal deformability of the star.
This relation is called the {\it adiabatic} approximation for the
tidal deformation of a star, which is valid only when the time scale
in the change of the weak tidal field is much longer than the
dynamical time scale of the star.  The tidal deformability is related
to the quadrupolar tidal Love number $k_{2}$ by
\begin{eqnarray}
\lambda = \frac{2}{3}R^{5}k_{2},
\end{eqnarray}
where $R$ is the radius of the (spherical) star in isolation.  For a
given EOS and a central density, one can calculate the quantities mass,
$R$, $k_{2}$, and $\lambda$ of neutron stars by solving the
Tolman-Oppenheimer-Volkoff equations and the metric perturbation
equations \cite{hinderer08,damour09}.

\subsection{The post-Newtonian description for the motion of a tidally interacting binary}

The motion of tidally interacting NS-NS binaries in close orbits is
affected by the stellar internal structure. As long as the degree of
the tidal interaction is small, the correction of this effect can be
described only through the tidal deformability
$\lambda$~\cite{vines11}.  The evolution of the orbital angular
velocity in the inspiral of a tidally interacting binary is described
by
\begin{eqnarray}
\frac{dx}{dt} = F(x,M_{A},M_{B},\lambda_{A},\lambda_{B}),
\end{eqnarray} 
where $x=(M \omega)^{2/3}$ with $\omega$ being the angular velocity
of the binary
and $M$ being the total ADM mass at the infinite separation. The
subscript (A or B) refers to each component of the binary. 
The function $F$ can be 
decomposed into the following two parts, 
\begin{eqnarray}
 F = F_{\rm{pp}}(x,M_{A},M_{B}) 
+ F_{\rm{tidal}}(x,M_{A},M_{B},\lambda_{A},\lambda_{B}),
\end{eqnarray}
where $F_{\rm{pp}}$ is the contribution of the point-particle part,
and $F_{\rm{tidal}}$ is the contribution associated with the tidal
interactions.  In this work, we adopt the Taylor T4 approximant for
the $F_{\rm{pp}}$ \cite{boyle07},
\begin{align}
& F_{\rm{pp}}^{\rm{T4}}  =  \frac{64\nu}{5M}x^{5}\left[1-\left(\frac{743}{336}+\frac{11}{4}\nu \right)x + 4\pi x^{3/2}\right.\\ \nonumber 
& + \left(\frac{34103}{18144}+\frac{13661}{2061}\nu + \frac{59}{18}\nu^{2}\right)x^{2} - \left(\frac{4159}{672}+\frac{189}{8}\nu \right)\pi x^{5/2} \\ \nonumber 
& + \left\{\frac{16447322263}{139708800}-\frac{1712}{105}\gamma -\frac{56198689}{217728}\nu \right.\\ \nonumber
&\left. + \frac{541}{896}\nu^{2}-\frac{5605}{2592}\nu^{3} + \frac{\pi^{2}}{48}\left(256 + 451\nu\right) -\frac{856}{105} \ln(16x) \right\}x^{3} \\ \nonumber
& + \left(-\frac{4415}{4032}+\frac{358675}{6048}\nu \left. + \frac{91495}{1512}\nu^{2}\right)\pi x^{7/2} \right],
\end{align}
where $\nu=M_{A}M_{B}/M^{2}$ is the symmetric mass ratio, and
$\gamma=0.577126...$ is an Euler's constant.  As shown by Boyle and
his collaborators~\cite{boyle07}, the evolution of the angular velocity of the
Taylor T4 approximant agrees well with those of the NR simulations for
the inspiral of equal-mass non-spinning binary black holes up to
$M\omega \sim 0.1$.  We adopt the tidal part which is derived
by Vines and his collaborators~\cite{vines11} as follows,
\begin{align}
&F_{\rm{tidal}} = \frac{32M_{A}\lambda_{B}}{5M^{7}}\left[12\left(1+11\frac{M_{A}}{M}\right)x^{10} \right.\\ \nonumber
&+ \big(\frac{4421}{28}-\frac{12263}{28}\frac{M_{B}}{M}+\frac{1893}{2}\left(\frac{M_{B}}{M}\right)^{2}\\ \nonumber
& \left. -661\left(\frac{M_{B}}{M}\right) ^{3} \big) x^{11} \right]
+ (A \leftrightarrow B).
\end{align}
In particular, for the case of equal-mass binaries, $F_{\rm{tidal}}$
is given by
\begin{eqnarray}
F_{\rm{tidal}} = \frac{52}{5M}x^{10} k_2 C^{-5}\left( 1 + \frac{5203}{4368}x\right),
\end{eqnarray}
where $C=M_{\rm{A}}/R_{\rm{A}}\left(=M_{\rm{B}}/R_{\rm{B}}\right)$ is
the compactness of the neutron star.  Although the tidal interaction
affects NS-NS inspirals only at 5PN order, its coefficient is of order
$10^4$ for typical neutron stars of radius 10 -- 15~km, $k_{2}\sim
0.1$, and $C \sim 0.14$ -- 0.20, and thus, it plays an important role
in the late inspiral stage. 

\subsection{Effective one body formalism for the motion of a tidally 
interacting binary}

The EOB formalism maps the dynamics of two point particles
to the Hamiltonian dynamics of an effective particle moving in an
effective external potential \cite{buonnano99,buonnano00,damour01}.
Because the EOB formalism goes beyond the adiabatic approximation for
binary inspirals, it is suitable for describing the late inspiral
stage of a binary, for which the adiabatic approximation is not very
accurate.  In this work, we employ the resummed EOB description which
is largely the same as that in Ref.~\cite{pan11} for the
point-particle part and as that in Ref.~\cite{damour12} for the tidal
part. 
 
The EOB effective metric is defined by
\begin{eqnarray}
ds_{\rm{eff}}^{2} = -A(r)dt^{2}+\frac{D(r)}{A(r)}dr^{2}+r^{2}\left(d\theta^{2}+\sin^{2}\theta d\phi^{2}\right),
\end{eqnarray}
where $(t, r,\phi)$ are the dimensionless coordinates and their canonical
momenta are $(p_{r},p_{\phi})$.  We replace the radial canonical
momentum $p_{r}$ with the canonical momentum $p_{r_{*}}$, where the
tortise-like radial coordinate $r_{*}$ is defined by
\begin{eqnarray}
\frac{dr_{*}}{dr}=\frac{\sqrt{D(r)}}{A(r)}.
\end{eqnarray}   
Then the binary dynamics can be described by the EOB Hamiltonian
\begin{eqnarray}
\hat{H}_{\rm{real}}(r,p_{r_{*}},p_{\phi}) = \frac{1}{\nu}\sqrt{1+2\nu \left(\hat{H}_{\rm{eff}}-1\right)}-\frac{1}{\nu} ,
\end{eqnarray}
where the effective Hamiltonian is defined by,
\begin{eqnarray} 
\hat{H}_{\rm{eff}} = \sqrt{p_{r_{*}}^{2}+A(r)\left(1+\frac{p_{\phi}^{2}}{r^{2}}+2\left(4-3\nu \right)\nu \frac{p_{r_{*}}^{4}}{r^{2}}\right)}.
\end{eqnarray}
The metric component $A(r)$ is decomposed into two parts as
\begin{eqnarray}
A(r) = A_{\rm{pp}}(r)+A_{\rm{tidal}}(r),
\end{eqnarray}
where $A_{\rm{pp}}(r)$ is the point-particle part and
$A_{\rm{tidal}}(r)$ is associated with the tidal effects.  The
point-particle part up to the 5PN order is given by
\begin{multline}
A_{\rm{pp}}(r)=P_{5}^{1}\Big[1-2u+2\nu u^{3}+a_{4}\nu u^{4}\\ 
+ a_{5} \nu u^{5} + a_{6}\nu u^{6} \Big],
\end{multline} 
where $a_{4}=94/3-41\pi^{2}/32$, $u = 1/r$, and $P^{1}_{5}$ denotes a
$(1,5)$ Pade approximant.  In the definition of $A(r)$, there are two
analytically undetermined parameters $(a_{5},a_{6})$, which correspond
to the 4PN and 5PN corrections.  Here, we adopt the values
$(a_{5},a_{6})=((-5.828-143.5\nu+477\nu^{2})\nu,184\nu)$ following
Ref.~\cite{pan11}.

The tidal part of $A(r)$ is given by
\begin{eqnarray}
A_{\rm{tidal}}(r) = \sum_{l \geq 2} -\kappa_{l}u^{2l+2}\hat{A}_{l}(u),
\end{eqnarray}
where $\hat{A}_{l}$ includes the PN tidal contributions for each
multipole, and $\kappa_{l}$ is its coefficient.  This coefficient is
related to the tidal Love number $k_{l}$ and the compactness
of two stars by
\begin{eqnarray}
\kappa_{l} = 2\frac{M_{B}M_{A}^{2l}}{M^{2l+1}}\frac{k^{A}_{l}}{C^{2l+1}_{A}} + (A \leftrightarrow B).
\end{eqnarray}
In this work, we include only the tidal-interaction part of the lowest
multipole $l=2$.  Up to the next-to-leading order, $\hat{A}_{l=2}$ is
given by~\cite{damour10}
\begin{eqnarray}
\hat{A}_{l=2}^{(\rm{A})}(u) = 1+\alpha_{1}^{(\rm{A})}u,
\end{eqnarray} 
where $\alpha_{1}^{(\rm{A})}=5M_{A}/2M$.  The tidal-interaction term
up to the 2PN corrections
is currently known, and the coefficient $\alpha_{2}$ is larger
than $\alpha_{1}$~\cite{bini12}.  In addition, an analysis in the
test-mass limit ($M_{\rm{A}}\ll M_{\rm{B}}$) suggests that the tidal
part $\hat{A}^{{\rm{(A)}}}_{l=2}$ has the pole-like behavior near the
last unstable orbit located at $3M_{\rm{B}}$ (the light ring orbit).  Thus, it
is reasonable to expect that the higher-PN corrections
would amplify the tidal effects.  In this work, we employ the resummed
version of the tidal metric including up to the
next-to-next-to-leading order given by
\cite{bini12}
\begin{eqnarray}
\hat{A}_{l=2}^{(\rm{A})}(u) = 1+\alpha_{1}^{(\rm{A})}u+\alpha_{2}^{(\rm{A})}
\frac{u^{2}}{1-\hat{r}_{\rm{LR}}u}, 
\end{eqnarray}
where 
\begin{eqnarray}
\alpha_{2}^{(\rm{A})}= 337M_{A}^{2}/28M^{2}+M_{A}/8M+3,
\end{eqnarray}
and
\begin{eqnarray}
\hat{r}_{\rm{LR}}(\nu,\kappa_{2}) = 3\left[ 1-\frac{5\nu}{3^{3}}+\frac{4}{3^{6}}\kappa_{2} + O(\nu^{2},\kappa_{2} \nu, \kappa^{2}_{2})\right], 
\end{eqnarray}
is the dimensionless radius of the light ring orbit. 

For the calculation of the binary orbit, we solve the EOB Hamilton
equations~\cite{buonnano99,buonnano00,damour01}
\begin{eqnarray}
&\displaystyle \frac{dr}{dt} = \frac{A(r)}{\sqrt{D^{0}_{3}(r)}}
\frac{\partial \hat{H}_{\rm{real}}}{\partial p_{r_{*}}},\\
&\displaystyle \frac{d\phi}{dt} = \frac{\partial \hat{H}_{\rm{real}}}
{\partial p_{\phi}},\\
&\displaystyle \frac{dp_{r_{*}}}{dt} = -\frac{A(r)}{\sqrt{D_{3}^{0}(r)}}
\frac{\partial \hat{H}_{\rm{real}}}{\partial r}+\hat{\mathcal{F}}_{\phi}
\frac{p_{r_{*}}}{p_{\phi}},\\
& \displaystyle \frac{dp_{\phi}}{dt} = \hat{\mathcal{F}}_{\phi},
\end{eqnarray}
where $D_{3}^{0}(r)$ is a $(0,3)$ Pade approximant of $D(r)$
\cite{pan11}, and 
$\hat{\mathcal{F}}_{\phi}$ is the radiation-reaction force given by
\begin{eqnarray}
\hat{\mathcal{F}}_{\phi} = -\frac{1}{8\pi \nu \hat{\omega}} \sum_{l=2}^{8}\sum_{m=1}^{l}
\left(m\hat{\omega} \right)^{2}|\frac{R}{M}h_{lm}|^{2}.
\end{eqnarray}
Here, $\hat{\omega} = d\phi/dt$, $h_{lm}$ denote the multipolar waveforms, and 
$R$ is the radius of extracting gravitational waves. 
The waveforms are described by
\begin{eqnarray}
h_{lm}=h_{lm}^{0}+h_{lm}^{\rm{tidal,A}}+h_{lm}^{\rm{tidal,B}},
\end{eqnarray}
where $h_{lm}^{0}$ denotes the inspiral and plunge waveform for a
binary black hole of mass $M_A$ and $M_B$, and $h_{lm}^{\rm{tidal,A}}$
is the contribution due to the tidal deformation of star A. In this
work, $h_{lm}^{0}$ is given by Eqs.~(13)--(22) of Ref.~\cite{pan11}
and $h_{lm}^{\rm{tidal,A}}$ is basically given by Eqs.~(A14)--(A17) of
Ref.~\cite{damour12}.  Because the 2PN term associated with the tidal
effect in the waveform is currently unknown, we include the
contributions of the tidal effect to $h_{lm}^{\rm{tidal,A}}$ up to the
1PN term even for the case that the next-to-next-to-leading order term
is taken into account in the radial potential. 

\begin{table*}[ht]
\caption[table]{Key parameters for the initial models adopted in the
  present numerical simulation. $M_{0}$ is the sum of the ADM masses
  of two neutron stars in isolation; $\nu$ is the symmetric mass
  ratio; $M_{0}^{\rm{ADM}}$ and $J^{\rm{ADM}}_{0}$ are the ADM mass
  and angular momentum of the system, respectively; $M_{*}$ is the
  baryon rest mass; $\omega_{0}$ is the angular velocity.  We also
  show the setup of the grid structure of our AMR algorithm.  $\Delta
  x =h_{6} =L/(2^{6}N)~(N=60)$ is the grid spacing for the
  highest-resolution domain with $L$ being the location of the outer
  boundaries for each axis.  $N$ specifies the grid size of the
  simulation with maximum $N=60$.  $\kappa_{2}$ is the parameter
  related to the tidal deformability.  $\omega_{\rm{contact}}$ is the
  orbital angular velocity at contact of the two neutron stars (see
  Sec. V. B). Here we use the unit $M_{\odot}=1$}\label{initial}
\begin{center}
 \begin{tabular}{lcccccccccc} \hline \hline
 \textrm{Model} & $M_{0}$& $\nu$ & $M^{\rm{ADM}}_{0}$ & $J^{\rm{ADM}}_{0}$ & $M_{*}$ &
$M\omega_{0}$ & $\Delta x/M$ & $\kappa_{2}$ & $M\omega_{\rm{contact}}$ & $N$ \\ \hline
APR4~~ & ~2.7~ & 0.25& ~2.68~ & ~7.65~ & ~3.00~ &~0.019~ & 0.0438 & 62.3 & 0.151 & (42,48,54,60)\\
H4 & 2.7 & 0.25 & 2.68 & 7.66 & 2.94 & 0.019 & 0.0560 & 215 & 0.112 & (42,48,54,60) \\
MS1 & 2.7 & 0.25 &2.68 &7.67 & 2.92 & 0.019 & 0.0595 &332 & 0.103 & (42,48,54,60) \\
APR4-1215 & ~2.7 & 0.24691358 & 2.68 & 7.28 & 3.01 & 0.0221 & 0.0438 & 65.8 & 0.151 & (40,50,60) \\
H4-1215 & ~2.7 & 0.24691358 & 2.68 & 7.56 & 2.94 & 0.019 & 0.0560 & 207& 0.114 &(40,50,60) \\ \hline \hline
\end{tabular}
\end{center}
\end{table*}

\begin{table*}[ht]
\caption[table]{Parameters of the piecewise-polytropic EOS, the
  compactness, and the tidal Love number of the neutron star of mass
  $1.35M_{\odot}$.}\label{eos}
\begin{center}
 \begin{tabular}{|c|cccccc|} \hline \hline
 EOS & $\log P_1 (\rm{dyne/cm}^{2}$) & $\Gamma_{1}$ & $\Gamma_{2}$ & $\Gamma_{3}$ &
$C(1.35)$ & $k_{2}(1.35)$\\
 \hline
APR4 & 34.269 & 2.830 & 3.445 & 3.348 & 0.179  & 0.091\\
H4 &34.669 & 2.909 & 2.246 & 2.144 & 0.146 & 0.115\\
MS1 & 34.858 & 3.224 & 3.033 & 1.325 & 0.138 & 0.133   
\\ \hline \hline
\end{tabular}
\end{center}
\end{table*}
  
\section{Numerical Relativity Simulation}

In this section, we briefly describe the formulation and the numerical
schemes of our NR simulation employed in this work. 

\subsection{Evolution and Initial Condition}

We follow the inspiral and merger of NS-NS binaries using our NR code,
called {\tt SACRA}, for which the details are described in
Ref.~\cite{yamamoto08}.  {\tt SACRA} employs a moving puncture version
of the Baumgarte-Shapiro-Shibata-Nakamura formalism~\cite{shibata95,
  baumgarte98,campanelli06} to solve Einstein's equations imposing the
equatorial symmetry (and $\pi$-symmetry for the equal-mass cases).  In the
numerical simulations, a fourth-order finite differencing scheme in
space and time is used implementing an adaptive mesh refinement
algorithm.  At refinement boundaries, a second-order interpolation
scheme is partly adopted.  The advection terms are evaluated by a
fourth-order non-centered finite differencing~\cite{brugmann08}.  A
fourth-order Runge-Kutta method is employed for the time evolution.
For the hydrodynamics, we employ a high-resolution central scheme
based on Kurganov and Tadmor scheme~\cite{kurganov00} with a
third-order piecewise parabolic interpolation and with a steep min-mod
limiter.

In this work, we prepare seven refinement levels both for efficiently
resolving two neutron stars by the finest-resolution domains and
for extracting gravitational waves in a local wave zone.  More precisely,
two sets of four finer domains comoving with each neutron star cover
the region of their vicinity.  The other three coarser domains cover
both neutron stars by a wider domain with their origins fixed
approximately at the center of the mass of the binary.  Each
refinement domain consists of the uniform, vertex-centered Cartesian
grids with $(2N+1,2N+1,N+1)$ grid points for $(x,y,z)$ with the
equatorial plane symmetry at $z=0$ imposed. The half of the edge
length of the largest domain (i.e., the distance from the origin to
outer boundaries along each axis) is denoted by $L$ which is chosen to
be larger than $\lambda_{0}$, where $\lambda_{0} = \pi/\omega_{0}$ is
the initial wave-length of gravitational waves and $\omega_{0}$ is the
initial orbital angular velocity. The grid spacing for
each domain is $h_{l}=L/(2^{l}N)$, where $l=0-6$.  In this work, we
choose $N=60$, 54, 48, and 42 for the resolution study.  With the
highest grid resolution, the semimajor diameter of each neutron star is
covered by about 100 grid points. 

We prepare NS-NS binaries in quasiequilibrium states for the initial
condition of numerical simulations by using a spectral-method library,
LORENE~\cite{lorene}.  To track more than 8 orbits, the orbital
angular velocity of the initial configuration is chosen to be $M\omega_{0}
= 0.019$ ($f=400$Hz for $M=2.7M_{\odot}$), where $M_{\odot}$ is the
solar mass.  The neutron stars are
assumed to have an irrotational velocity field, which is believe to be
an astrophysically realistic 
configuration~\cite{bildsten92,kochaneck92}.  The parameters for the
initial models are listed in Table~\ref{initial}. 

\subsection{Equation of State}

In this work, we employ a parameterized piecewise-polytropic EOS
proposed by Read and her collaborators~\cite{read09}. This EOS is written in
terms of four segments of polytropes
\begin{align}
P = & K_{i}\rho^{\Gamma_{i}} \\ \nonumber
&\text{( for $\rho_{i}\leq \rho<\rho_{i+1}$, $0\leq i \leq 3$)},
\end{align}
where $\rho$ is the rest-mass density, $P$ is the pressure, $K_{i}$ is
a polytropic constant, and $\Gamma_{i}$ is an adiabatic index.  
At each boundary of the piecewise
polytropes, $\rho=\rho_{i}$, the pressure is required to be
continuous, i.e.,
$K_{i}\rho_{i}^{\Gamma_{i}}=K_{i+1}\rho_{i}^{\Gamma_{i+1}}$.  Read
and her collaborators determine these parameters in the following
manner~\cite{read09}.  First, they fix the EOS of the crust as
$\Gamma_{0}=1.357$ and $K_{0}=3.594\times 10^{13}$ in cgs units.
Then they determine $\rho_{2}=1.85\rho_{\rm{nucl}}$ and
$\rho_{3}=3.70\rho_{\rm{nucl}}$ where $\rho_{\rm{nucl}}=2.7\times
10^{14}$ $\rm{g/cm^{3}}$ is the nuclear saturation density.  With this
preparation, they choose the following four parameters as a set of
free parameters: $\{P_{1}, \Gamma_{1}, \Gamma_{2}, \Gamma_{3}\}$.
Here $P_{1}$ is the pressure at $\rho=\rho_{2}$, and from this,
$K_{1}$ and $K_{i}$ are determined by
$K_{1}=P_{1}/\rho_{2}^{\Gamma_{1}}$ and
$K_{i+1}=K_{i}\rho_{i}^{\Gamma_{i}-\Gamma_{i+1}}$.  Therefore the EOS
is specified by choosing the four parameters $\{P_{1}, \Gamma_{1},
\Gamma_{2}, \Gamma_{3}\}$.  In this work, we choose the three sets of
piecewise-polytropic EOS as listed in Table~\ref{eos}.

We describe the low-density part of the EOS only with a single
polytrope, because the elastic property of the crust yields a very
small correction to the tidal number \cite{penner11}.  Thus, our
approximate treatment for the low-density part is acceptable.

\subsection{Extraction of Gravitational waves}

Gravitational waves are extracted by calculating the complex Weyl
scalar $\Psi_{4}$ ~\cite{yamamoto08} from which
gravitational waveforms are determined by
\begin{eqnarray}
h_{+}(t)-ih_{\times}(t) = -\lim_{r\rightarrow \infty}
\int^{t}dt^{\prime}\int^{t^{\prime}}dt^{\prime \prime}
\Psi_{4}(t^{\prime \prime},r).
\end{eqnarray}
Here we omit arguments $\theta$ and $\phi$.  In the spherical
coordinate $(r, \theta, \phi)$, $\Psi_{4}$ can be expanded in the form
\begin{eqnarray}
\Psi_{4}(t, r, \theta, \phi) = \sum_{lm}\Psi_{4}^{lm}(t,r)_{-2}
Y_{lm}(\theta, \phi),
\end{eqnarray}
where $_{-2}Y_{lm}$ are spin-weighted spherical harmonics of weight
$-2$ and $\Psi_{4}^{lm}$ are expansion coefficients defined by this
equation.  In this work, we focus only on the $(l,|m|)=(2,2)$ mode.
The gravitational-wave phase $\phi_{\rm{NR}}$ is defined by
\begin{eqnarray}
\Psi_{4}^{22}(t,r) = A_{22}(t,r)e^{i\phi_{\rm{NR}}(t,r)},
\end{eqnarray}
where $A_{22}$ denotes the amplitude and it is real.  We evaluate
$\Psi_{4}$ at a finite spherical-coordinate radius $r/M_{\odot}=200$,
240, 300,~and~$400$.  To compare the waveforms extracted at different
radii, we use the retarded time defined by
\begin{eqnarray}
t_{\rm{ret}} = t - r_{*}.
\end{eqnarray}
Here, $r_{*}$ is the tortoise coordinate defined by
\begin{eqnarray}
r_{*} = r_{\rm{A}} + 2M\ln \left(\frac{r_{\rm{A}}}{2M}-1\right),
\end{eqnarray}
where $r_{\rm{A}}=\sqrt{A/4\pi}$ with $A$ being the proper surface area of
the extraction sphere.

\begin{figure*}[ht]
 \begin{tabular}{l l}
 \rotatebox{0}{\includegraphics[scale=0.65]{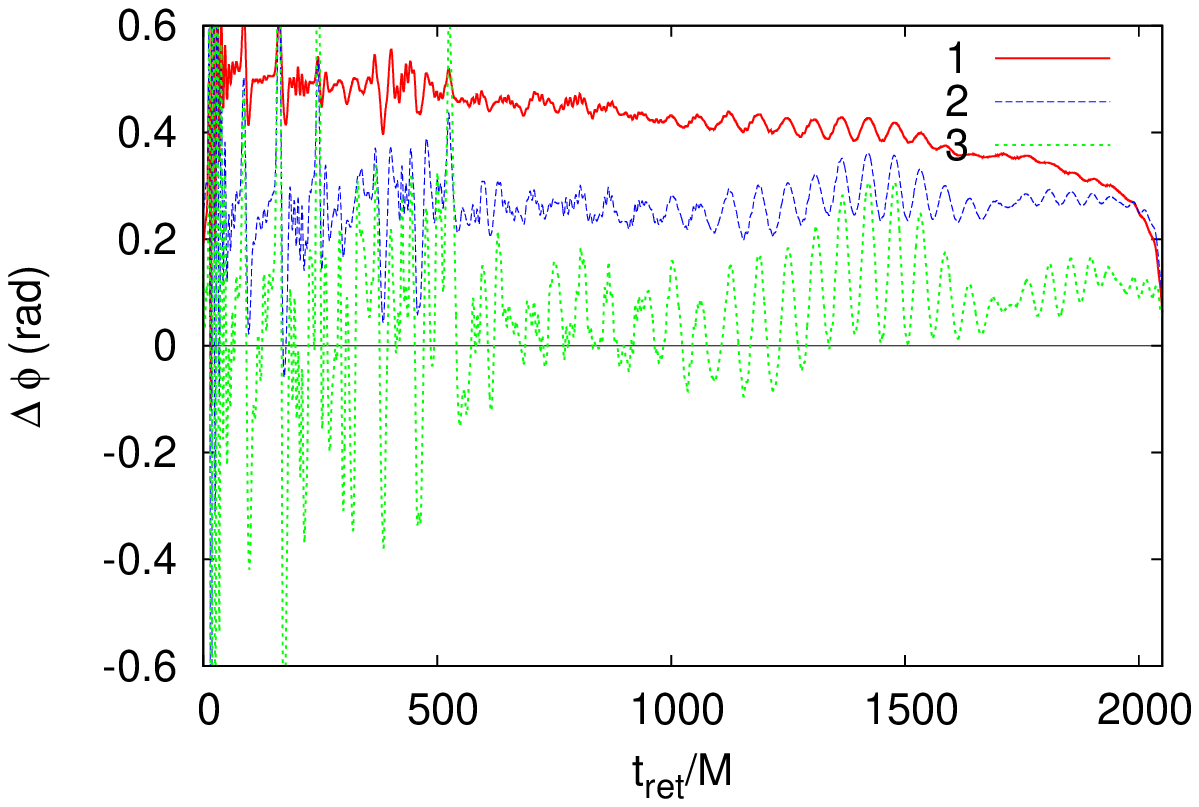}}
 \rotatebox{0}{\includegraphics[scale=0.65]{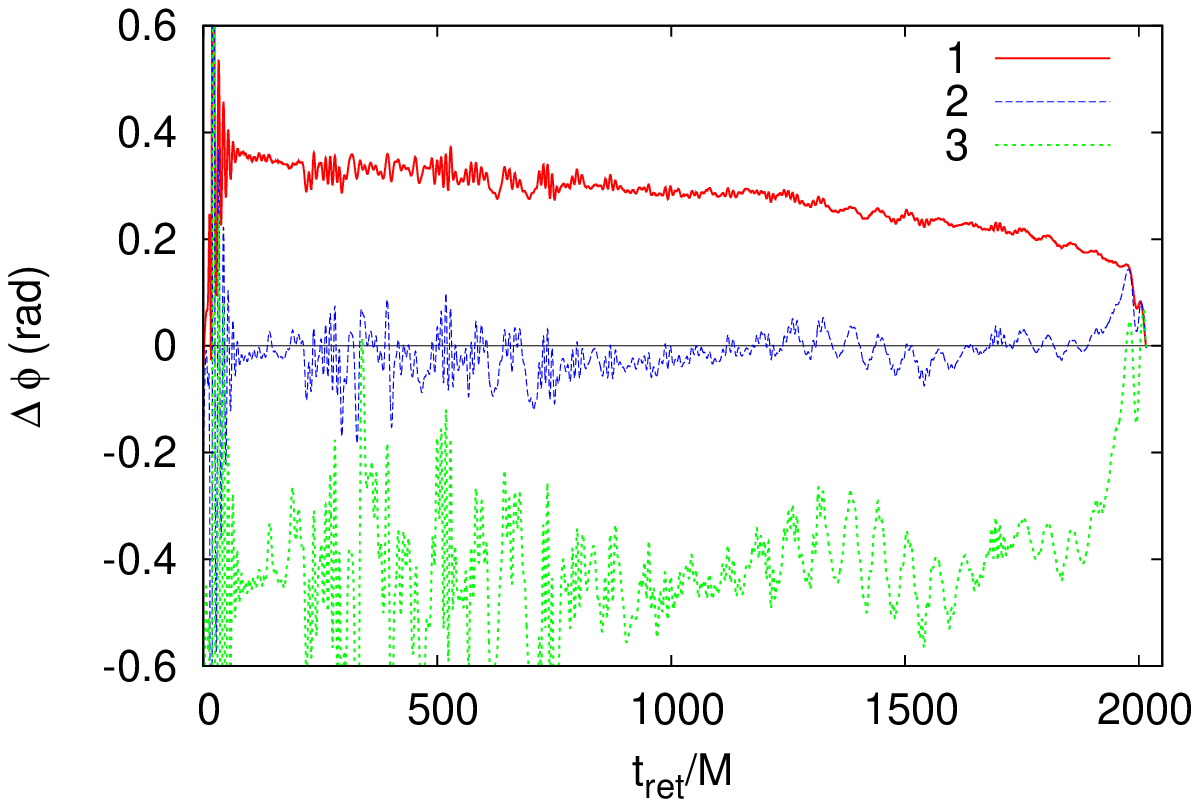}}\\
 \end{tabular}
 \caption{The phase errors for APR4 (left) and H4 (right).  The curve
   1 denotes the difference between the gravitational-wave phase
   extracted at $r/M_{\odot}=400$ and an extrapolated
   gravitational-wave phase obtained using the data at
   $r/M_{\odot}=300$ and $400$.  The curve 2 denotes the difference
   between the extrapolated gravitational-wave phase obtained using
   the data at $r/M_{\odot}=300~\rm{and}~400$ and that obtained using
   the data at $r/M_{\odot}=240,~300,$ and~$400$.  The curve 3 denotes
   the difference between the extrapolated gravitational-wave phase
   obtained using the data at $r/M_{\odot}=240,~300,~\rm{and}~400$ and
   that obtained using the data at
   $r/M_{\odot}=200,~240,~300,~\rm{and}~400$.  }
 \label{radii}
\end{figure*}

\section{Data Analysis of Numerical-Relativity Simulations}

The extrapolation with respect to the extraction radius and grid
resolution plays a key role to obtain the physical waveforms in NS-NS
inspirals from the results of NR simulations. Here, we focus in
particular on deriving the accurate gravitational-wave phase by
extrapolation because it carries the most important information in the
matched filtering for data analysis.  Thus, the goal of this section
is to construct the extrapolated gravitational-wave phase. 

\subsection{Extrapolation to infinity}

Because gravitational waves are extracted at finite radii, the
extracted waveform does not fully agree with the waveform at infinity.
For obtaining the hypothetical gravitational-wave phase at infinity,
we first need to estimate the error due to the finite-radii extraction
and then to extrapolate the gravitational-wave phase to infinity.  For
this purpose, we assume that the gravitational-wave phase is described
by a polynomial~\cite{boyle07},
\begin{eqnarray}
\phi(t_{\rm{ret}},r) = \phi^{(0)}(t_{\rm{ret}}) + \sum_{i=1}^{s} 
\frac{\phi^{(i)}(t_{\rm{ret}})}{r^{i}}. 
\end{eqnarray}
Here, $\phi^{(0)}(t_{\rm{ret}})$ is considered to be the
gravitational-wave phase extrapolated at $r \rightarrow \infty$,
and $(s+1)$ is the number of extraction radii used. We determine it
by extrapolating the gravitational-wave phases extracted at
$r/M_{\odot}=200,~240,~300,~\rm{and}~400$.

Figure \ref{radii} shows the differences among the hypothetically
extrapolated gravitational-wave phases at infinity obtained by
different numbers of extrapolation radius, $s$.  The curves labeled by 1 --
$3(=s-1)$ in Fig.~\ref{radii} denote the differences between a
radius-extrapolated gravitational-wave phase with the ($s-1$)-th order
polynomial and that with the $s$-th order one.  Here, we set the
phase difference to be zero at the merger.  Note that the difference
between the ($s-1$)-th and $s$-th order extrapolated
gravitational-wave phases accumulates mainly just after the contact of the two stars and
its value is less than about 0.5~radian. In this paper, we do not pay
attention to the gravitational-wave phase after contact of the two stars.
As shown in the next
subsection, furthermore, this accumulated gravitational-wave phase
difference of the radius extrapolation is much smaller than the
magnitude of other error sources. Therefore, in this paper, we neglect
the phase error associated with the finite-radius extraction, and use
the resolution-extrapolated gravitational-wave phase with the
extracted radius $r/M_{\odot}=400$ as the extrapolated wave phase.  We
note however that in a high-resolution study, the error associated
with the finite-radius extraction could be comparable to or smaller
than the error due to the finite grid resolution.

\begin{figure*}[htbp]
 \begin{tabular}{l l}
 \rotatebox{0}{\includegraphics[scale=0.65]{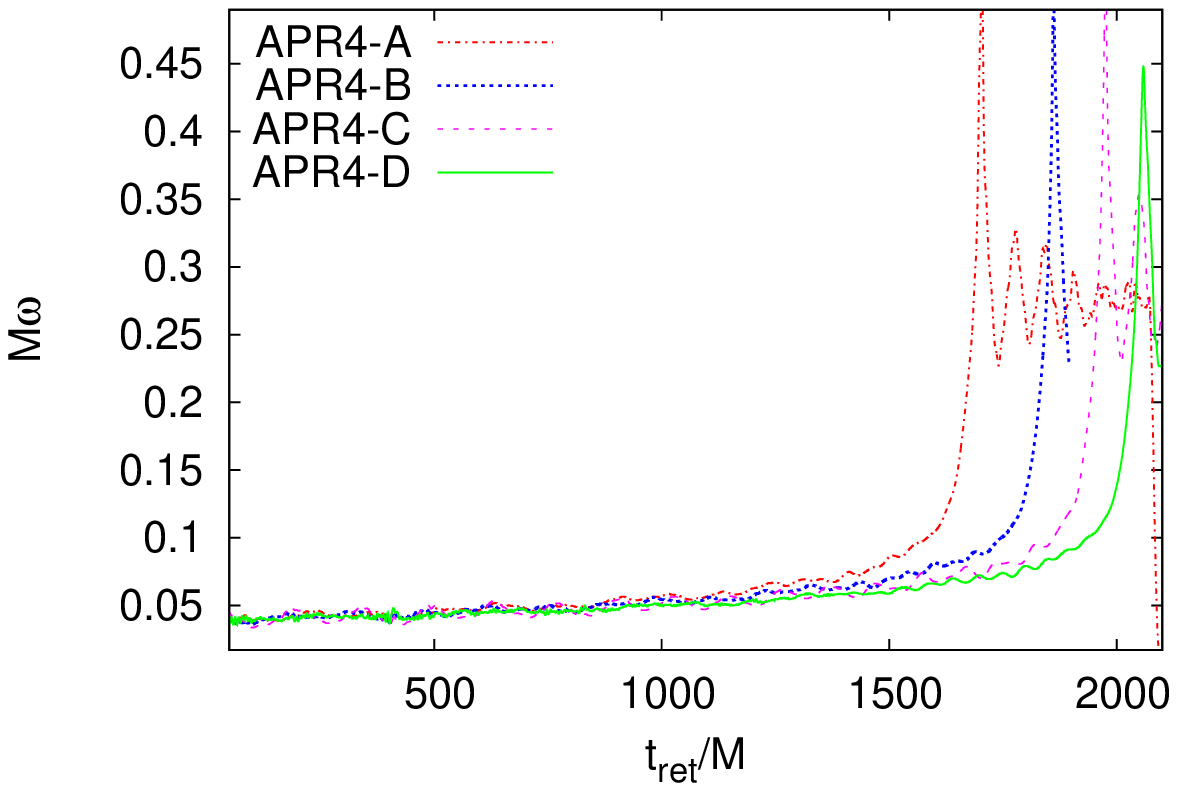}}
 \rotatebox{0}{\includegraphics[scale=0.65]{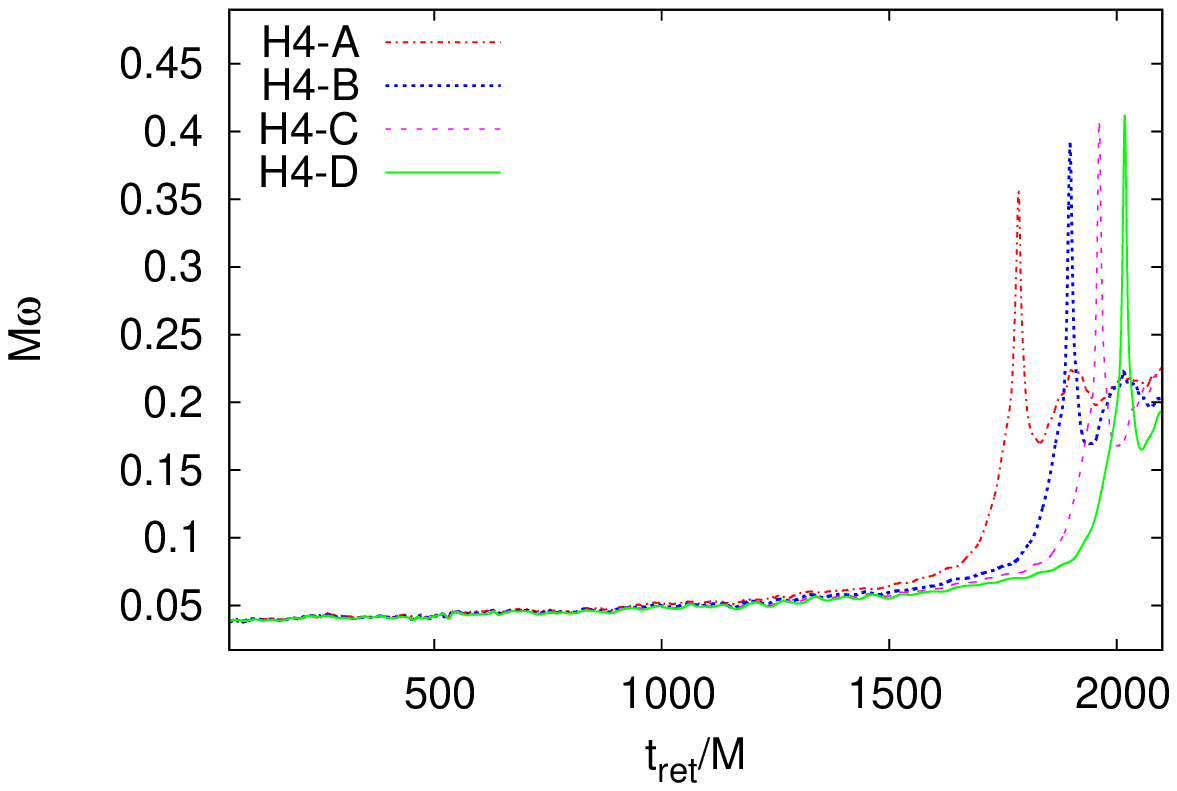}}\\
 \end{tabular}
 \caption{The evolution of gravitational-wave angular velocity for
   four different resolutions for APR4 (left panel) and H4 (right
   panel).  The curves with marks A, B, C, and D denote the results
   for the simulations with $N = (42, 48, 54, 60)$, respectively.  }
 \label{mw1}
\end{figure*}
\begin{figure*}[ht]
\begin{center}
\centerline{
\begin{tabular}{l l}
\rotatebox{0}{\includegraphics[scale=0.65]{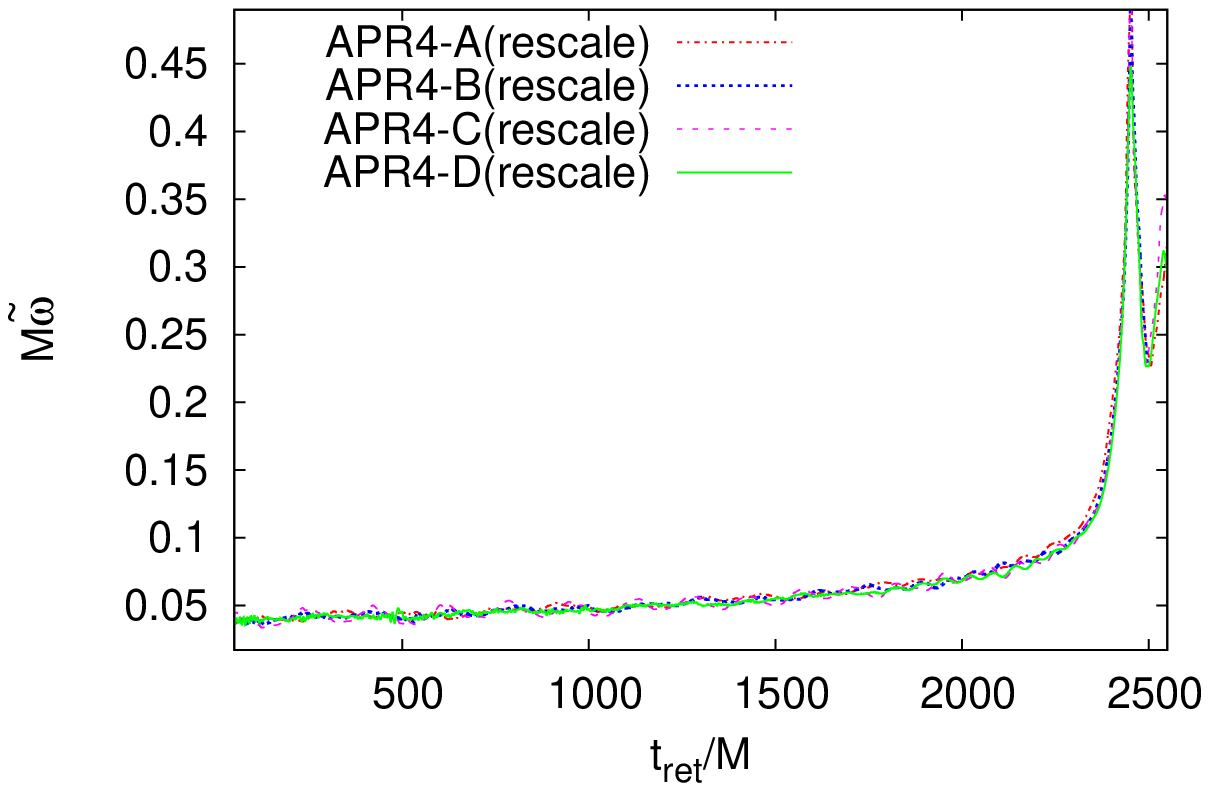}}
\rotatebox{0}{\includegraphics[scale=0.65]{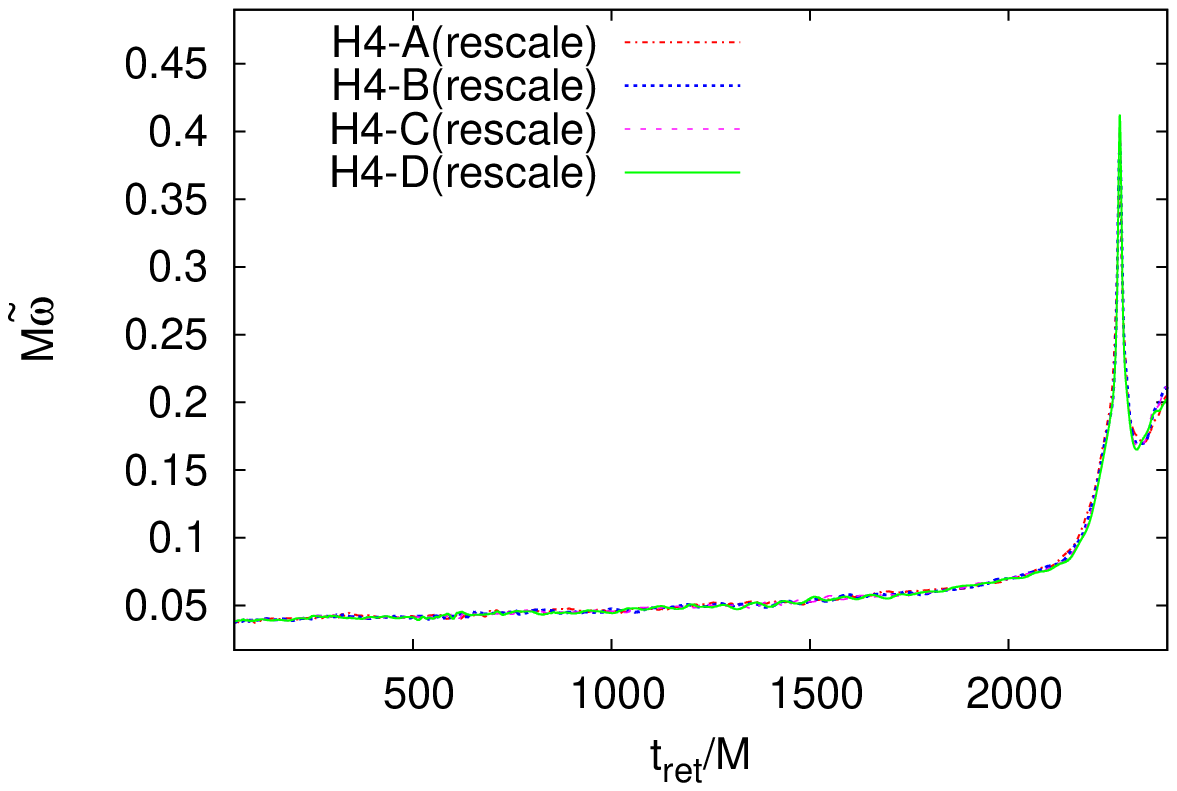}}\\
\rotatebox{0}{\includegraphics[scale=0.65]{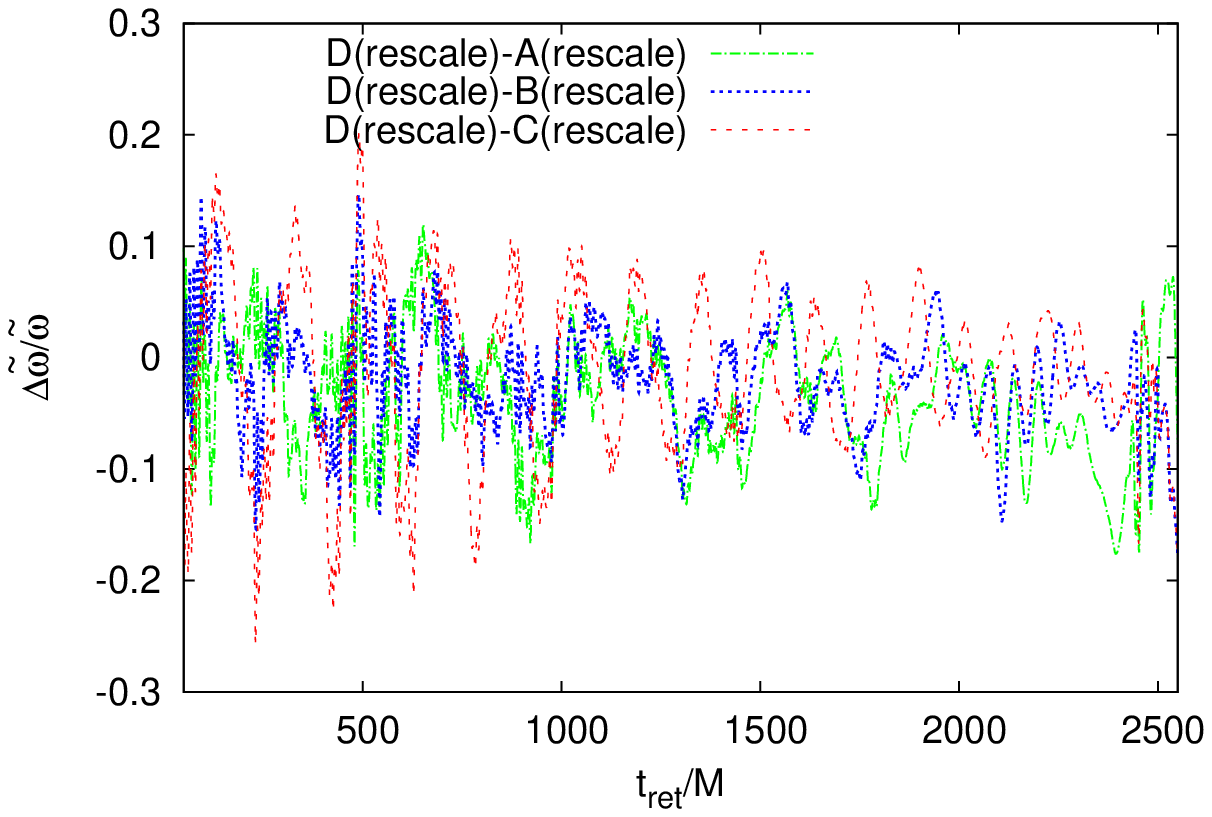}}
\rotatebox{0}{\includegraphics[scale=0.65]{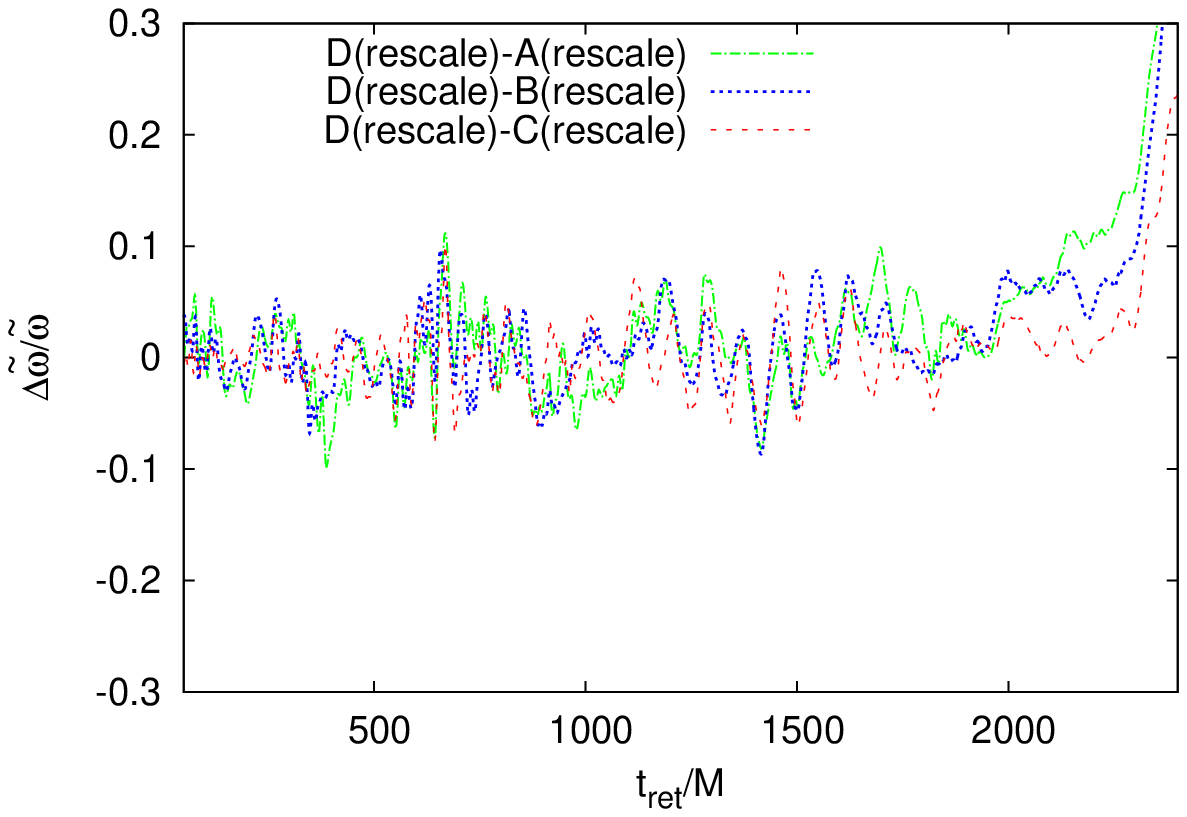}}\\
\end{tabular}
}
\caption{Convergence of the rescaled gravitational-wave angular
  velocity.  {\it Upper panels}: the evolution curves of the rescaled
  gravitational-wave angular velocity for APR4 (left panel) and H4
  (right panel).  {\it Lower panels}: the difference in the rescaled
  gravitational-wave frequency between different resolutions for APR4
  (left panel) and H4 (right panel). Here we choose $n=1.8$.}

\label{mw2}
\end{center}
\end{figure*}

\subsection{Extrapolation to infinite resolution}
A NR waveform is derived as a solution of discretized Einstein's
equations and hydrodynamics equations.  Associated with the finite
differencing, a truncation error is yielded.  For the purpose of comparing the
NR waveform with the PN and EOB waveforms, we have to obtain a
hypothetically physical waveform by extrapolating the NR waveform
obtained in the finite grid resolution for the limit of the infinite
grid resolution.  The order of the convergence of our NR simulation
with respect to the grid spacing is found to be about $1.8$. Because
the order of the convergence has an uncertainty, we conservatively
assume that the order of the convergence is $1.8 \pm 0.2$ in 
the following. 

For understanding the convergence property of the numerical results,
we check the convergence for the evolution of the gravitational-wave
frequency performing simulations with four different grid resolutions
(see Fig.~\ref{mw1} for gravitational-wave angular velocity). Here, we
refer to the data for $N=(42, 48, 54, 60)$ as (A, B, C, D).  Although
the gravitational-wave frequencies agree with each other at
$t_{\rm{ret}}=0$, their subsequent evolution disagrees with each other.
Obviously, the gravitational-wave frequency in the lower grid
resolution evolves more rapidly than that in the higher grid resolution.
The rapid evolution for the lower-resolution simulations may be
ascribed to larger numerical dissipation of the angular momentum.

However, one can find the similarity among four curves of the
gravitational-wave frequency for the different grid resolutions and this
enables us to obtain an extrapolated waveform.  To show this
similarity, we normalize the time variable by using the time at the
onset of the merger, which is defined by $t_{m}(\Delta x) =
t|_{\omega_{\rm{max}}}(\Delta x)$:
\begin{eqnarray}
t \rightarrow \tilde{t}(t) = \frac{t}{\Lambda(\Delta x)}.
\end{eqnarray}
Here, $\Lambda(\Delta x) = t_{m}(0)/t_{m}(\Delta x)$, $\Delta x$ is
the grid spacing of the simulation ($\propto N^{-1}$), and $t_{m}(0)$
is the extrapolated merger time.  For obtaining the extrapolated merger time, we assume
that the merger time as a function of grid resolutions is described by a binomial
\begin{eqnarray}
t_{m}(\Delta x) = t_{m}(0) + K(\Delta x)^{n},
\end{eqnarray}
where $t_{m}(0)$, $n$, and $K$ are constants which are determined by
the least-square fitting method.  We find that the best fitted value
of $n$ is $\approx 1.8$ with a dispersion $\sim 0.2$, and thus, we set
the order of the convergence $n$ is to be $1.8 \pm 0.2$.  The
resulting fitted values for $n=(1.6,1.8,2.0)$ are
$t_{m}(0)=(2521M,2452M,2397M)$ for APR4, 
$t_{m}(0)=(2391M,2274M,2238M)$ for H4, and $t_{m}=(2145M,2145M,2091M)$
for MS1.

We define the rescaled gravitational-wave frequency by
\begin{eqnarray}
\tilde{\omega}(t,\Delta x)\equiv \omega(\tilde{t}(t),\Delta x). \label{rescale_w}
\end{eqnarray}
The evolution curves of the rescaled gravitational-wave frequency
$\tilde{\omega}$ for the different grid resolutions agree with each other,
as shown in the upper two panels of
Fig.~\ref{mw2} (here, we set $n=1.8$). The differences among the
rescaled gravitational-wave frequencies of the different grid
resolutions are within $\sim 10\%$ for APR4 and $\sim 5\%$ for H4, as
shown in the lower panels of Fig.~\ref{mw2}.  Therefore, we conclude
that the numerical results for the gravitational-wave frequency
(angular velocity) is written in the form
\begin{eqnarray}
\tilde{\omega}(t,\Delta x) = \tilde{\omega}(t,0)+\omega_{\rm{r}}(t,\Delta x),\label{eqw}
\end{eqnarray}
where $\omega_{\rm{r}}(t,\Delta x)$
is a function that depends on the grid resolution,
which should have been eliminated systematically by extrapolation.  
However, because the value of $\omega_{\rm{r}}(t,\Delta x)$ randomly
fluctuates, it cannot be eliminated by extrapolation at each
given moment.
Thus, the extrapolated gravitational-wave frequency
can be only approximately obtained by simply rescaling the time variable of
the simulations. 

\begin{figure}[t]
 \includegraphics[width=80mm,clip]{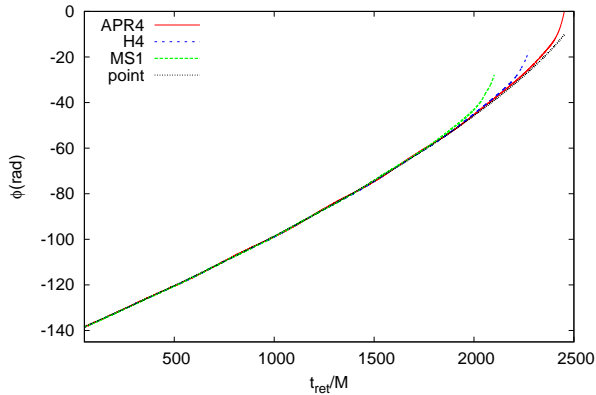}\\
 \caption{The extrapolated gravitational-wave phase for the three EOSs.
    Here, we set the order of the convergence to be 1.8.}
 \label{ph_eos}
\end{figure}

We proceed to extrapolate the rescaled gravitational-wave phase, which
will be compared with the gravitational-wave phases derived in the PN
and EOB approaches in the next section.
We define the rescaled gravitational-wave phase as
\begin{eqnarray}
\tilde{\phi}(t,\Delta x)\equiv \int_{0}^{t}\tilde{\omega}(t',\Delta x)dt'.\label{rescale_p}
\end{eqnarray}
Substituting Eq.~(\ref{rescale_w}) into this equation yields
\begin{eqnarray}
\tilde{\phi}(t,\Delta x)=\Lambda(\Delta x)\phi(\tilde{t},\Delta x),
\end{eqnarray}
and combining Eq.~(\ref{eqw}) and Eq.~(\ref{rescale_p}) yields
\begin{eqnarray}
\tilde{\phi}(t,\Delta x)=\tilde{\phi}(t,0)+\int_{0}^{t}\epsilon(t',\Delta x)dt',\label{rescale_p2}
\end{eqnarray}
where $\tilde{\phi}(t,0)$ is the extrapolated gravitational-wave phase.
Assuming the second term in the right-hand side of this equation can be described as 
$\phi_{\rm{r}}(t)\Delta x^{n}$, we obtain
\begin{eqnarray}
\Lambda (\Delta x)\phi(\tilde{t},\Delta x) = \tilde{\phi}(t,0) + \phi_{\rm{r}}(t)\Delta x^{n}.\label{eqp}
\end{eqnarray}
Here $\tilde{\phi}(t,0)$ and $\phi_{\rm{r}}(t)$ are
functions which are determined by the least-square fitting
of the numerical data $\phi_{\rm{NR}}(\tilde{t},\Delta x)$.
We again set the order of the convergence $n$ to be $1.8 \pm 0.2$. 
Figure~\ref{ph_eos}
shows the evolution curves of the extrapolated gravitational-wave phase
for the different EOS.

\begin{figure*}[ht]
 \begin{tabular}{l l}
\rotatebox{0}{\includegraphics[scale =0.65]{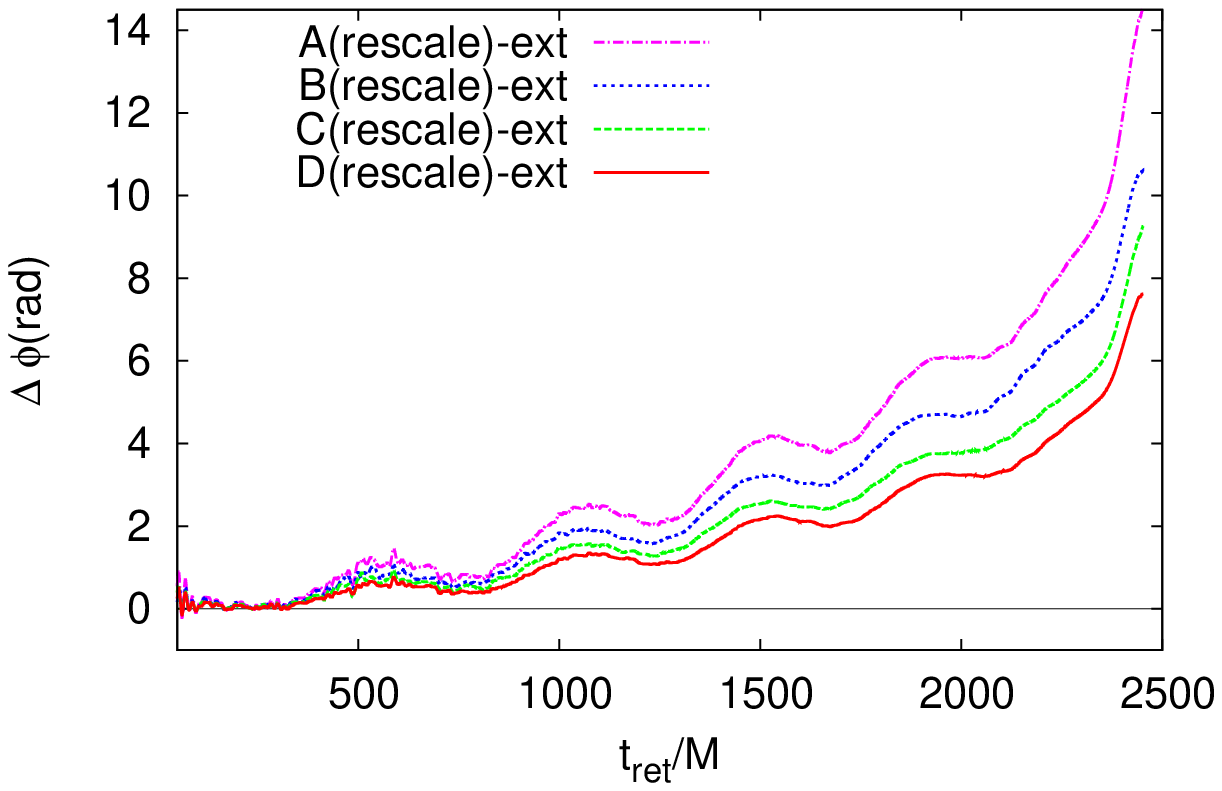}}
\rotatebox{0}{\includegraphics[scale=0.65]{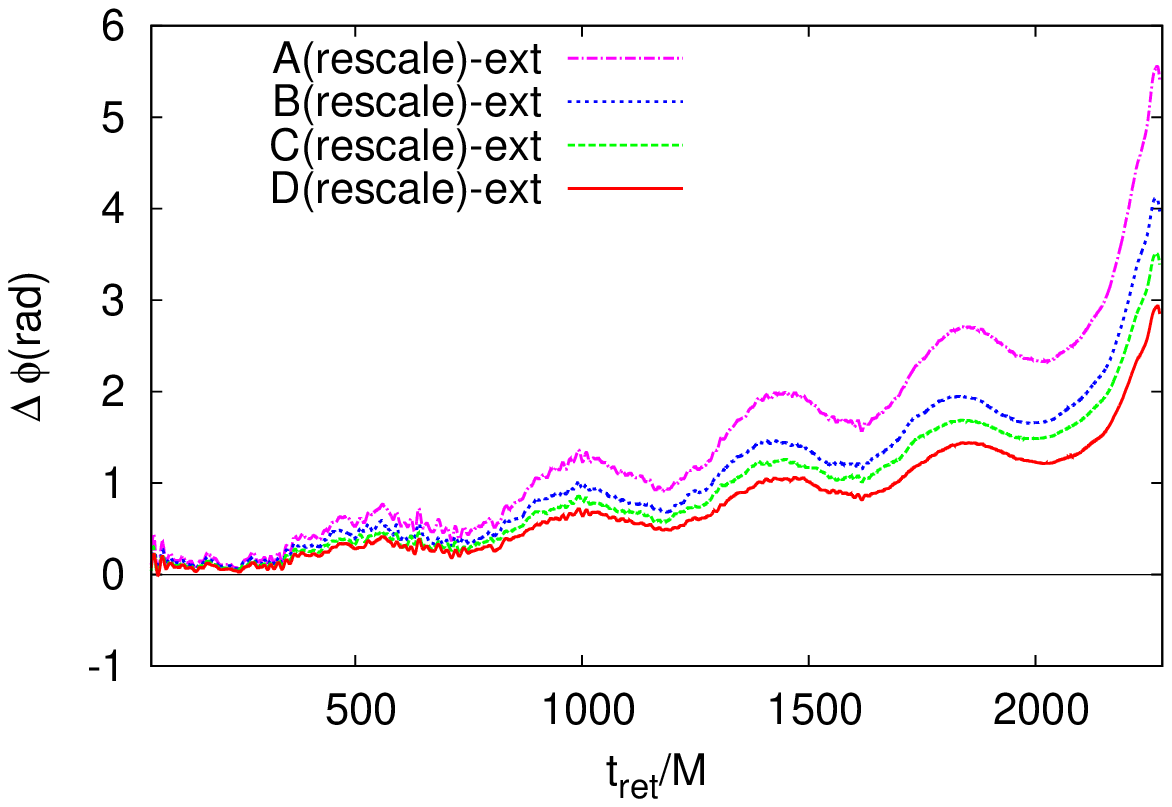}}\\
\rotatebox{0}{\includegraphics[scale=0.65]{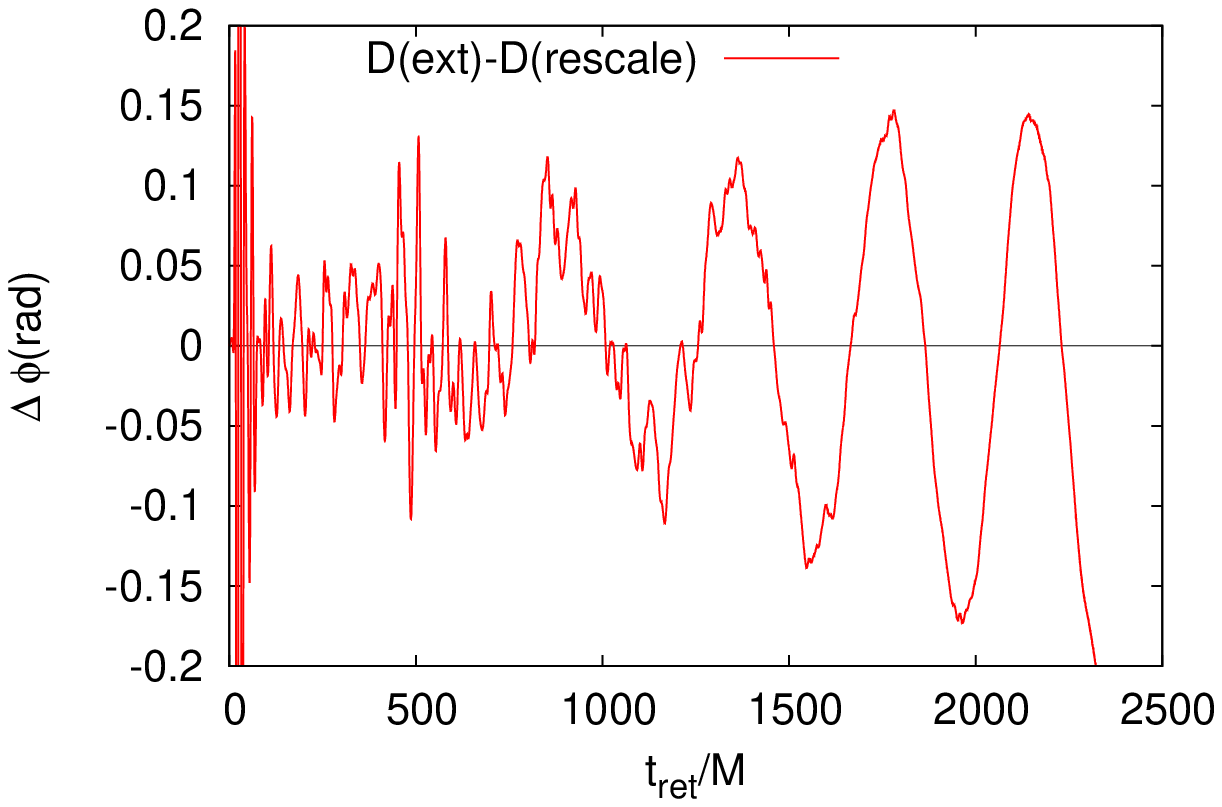}}
\rotatebox{0}{\includegraphics[scale=0.65]{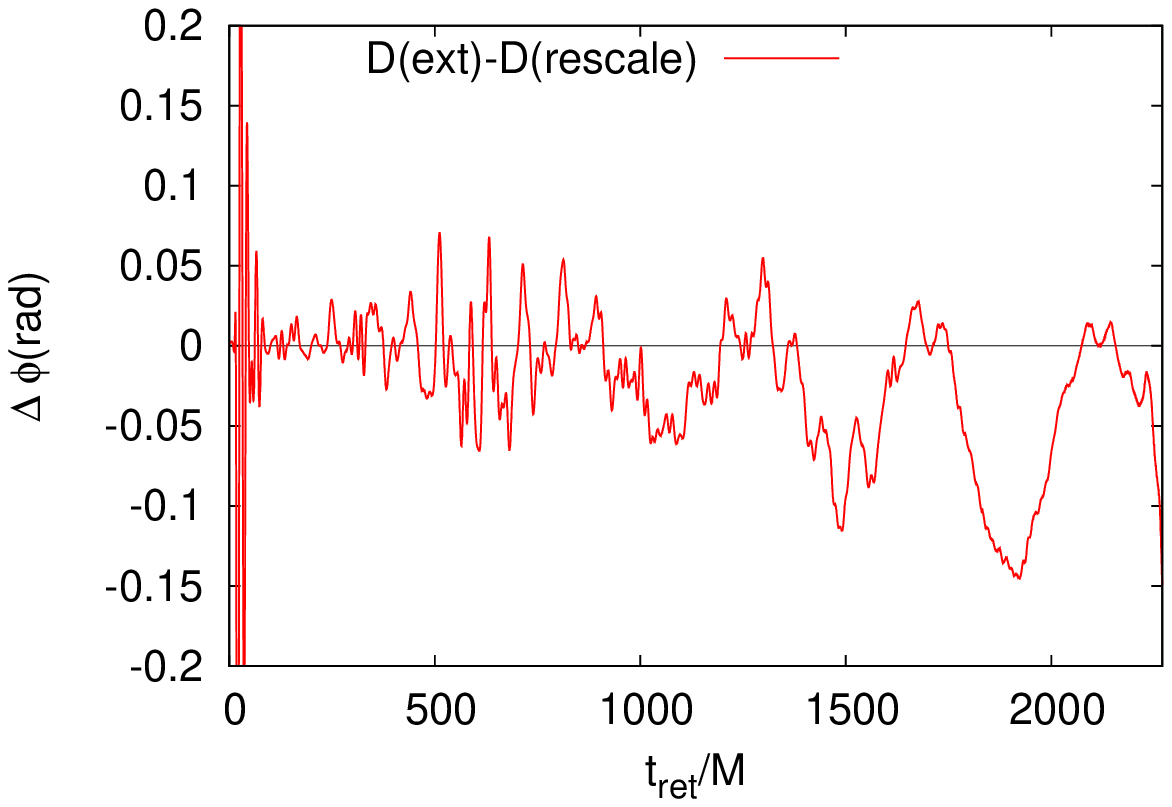}}\\
 \end{tabular}
 \caption{Convergence of the gravitational-wave phase for APR4 (left)
   and H4 (right).  In the upper panels, the curves show the phase
   differences between the extrapolated gravitational-wave phase and
   the rescaled numerical results. Here, we set the order of the
   convergence $n$ to be 1.8 for deriving the extrapolated
   gravitational-wave phase. In the lower panels, the curves show the
   difference between the rescaled gravitational-wave phase in the resolution D
   and the extrapolated gravitational-wave phase which is obtained by
   extrapolating the rescaled gravitational-wave phases of the resolution A, B,
   and C to the resolution D.  }
 \label{ph}
\end{figure*}

The differences in the rescaled gravitational-wave phase $\tilde{\phi}_{\rm{NR}}(t,\Delta x)$
obtained for runs with the
different grid resolutions and the extrapolated gravitational-wave
phase $\tilde{\phi}(t,0)$ for $n=1.8$ are shown in Fig.~\ref{ph}.  The accumulated
difference between the extrapolated gravitational-wave phase and that
of the highest grid resolution at the onset of the merger is $\sim
8$~radian for APR4 and $\sim 3$~radian for H4. These values imply
that the simulation for the neutron star which has the larger radius is relatively
well convergent. These phase differences are much
larger than the possible error in the gravitational-wave phase
associated with the finite-radius extraction, $\lesssim 0.5$~radian.
Thus, we conclude that the resolution extrapolation of the
gravitational-wave phase is essential to construct the extrapolated
physical waveform for our current NR simulations. 
The lower panels in Fig.~\ref{ph} show the
difference between the gravitational-wave phase of the resolution D,
$\tilde{\phi}_{\rm{NR}}(t,\Delta x_{\rm{D}})$, and the one which is obtained by
extrapolating the gravitational-wave phase of the resolutions
A, B, and C to the resolution D, $\tilde{\phi}(t,\Delta x_{\rm{D}})$.
The difference $\Delta \phi$ is within 0.15 radian up to the merger.
Here, this error is caused mainly by a modulation associated with
an unphysical orbital eccentricity.
The upper panels of Fig.~\ref{evolv-w} show the evolution curves of
the extrapolated gravitational-wave phase for the different choice of $n$.

For the brief check of the validity of the extrapolation,
we calculate the dispersion $\sigma(t)$ of the extrapolation as
\begin{multline}
\sigma(t)^{2} =\frac{1}{N_{\rm{R}}}\sum_{i=1}^{N_{\rm{R}}}\left( \tilde{\phi}(t,\Delta x_{i})-\tilde{\phi}_{\rm{NR}}(t,\Delta x_{i})\right)^{2}, 
\end{multline}
where $\tilde{\phi}_{\rm{NR}}(\tilde{t},\Delta x_{i})$ is the rescaled gravitational-wave phase of
the numerical data, $i=1\sim N_{\rm{R}}$ denote the numerical run with the different grid resolutions,
e.g.~$N=(42,48,54,60)$,
and $N_{\rm{R}}$ is the total number of them, e.g.~$N_{\rm{R}}=4$.
As shown later, the value of the dispersion is in the range 0.01~radian (in the early part)
to 0.4~radian (just before the merger).
We regard this dispersion as an error due to the resolution extrapolation.
Hereafter we use only the extrapolated quantities and
omit the tilde of them.

\section{Comparison of the NR wave phase with the PN and EOB wave phases}

\subsection{Matching Procedure}

In the analysis, we first match an early part of the
gravitational-wave phase obtained by extrapolating NR results ,
$\phi_{\rm{NR}}(t)$, with that of the analytic ones,
$\phi_{\rm{PN}}(t)$ (the wave phase derived in the PN calculation) 
and $\phi_{\rm{EOB}}(t)$ (the wave phase derived in the EOB calculation) , 
by minimizing the following quantity, 
\begin{multline}
\int_{t_{1}}^{t_{2}} \left( \Delta \phi(t)\right)^{2} dt \\= \int_{t_{1}}^{t_{2}} \left(\phi_{\rm{NR}}(t)-\phi_{\rm{PN/EOB}}(t-t_{\rm{s}})-\phi_{\rm{s}}\right)^{2}dt,
\end{multline} 
where $\phi_{\rm{s}}$ and $t_{\rm{s}}$ are the fitting parameters.
The initial and final time of the integral, $t_{1}$ and $t_{2}$, are
chosen as follows.  As shown in the previous section, the curve of the
extrapolated gravitational-wave phase obtained from the results of NR
simulations has a small modulation due to the orbital eccentricity.
Thus, $(t_{1},t_{2})$ are chosen to cover a range between two adjacent
local maxima or local minima of the curve in an early part of the
gravitational-wave phase.  This choice of the matching region allows
us to match the gravitational-wave phase numerically obtained with the
PN/EOB phases smoothly using the least-square fitting.  Here, we
choose the two adjacent local minima as the matching boundaries as
follows: $(t_{1},t_{2})=(583M,1030M)$ for APR4 (which corresponds to
$M\omega \in [0.042:0.048]$), $(t_{1},t_{2})=(550M,1000M)$ for H4
(which corresponds to $M\omega \in [0.042:0.046]$), and
$(t_{1},t_{2})=(550M,960M)$ for MS1.

\subsection{Comparison}

\begin{figure*}[ht]
 \begin{tabular}{l l}
\hspace{-1.0cm} \rotatebox{0}{\includegraphics[scale =0.7]{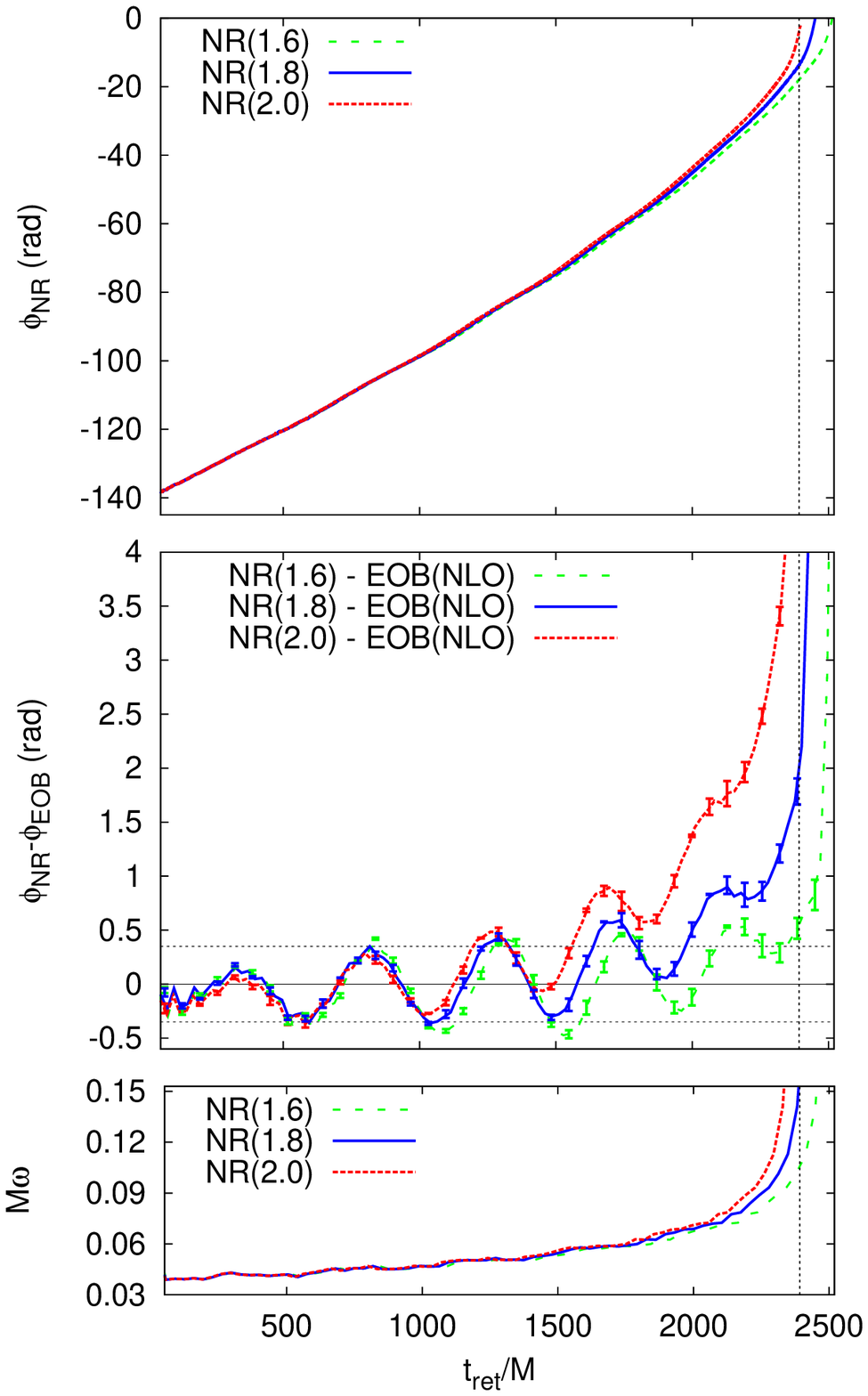}}
 \rotatebox{0}{\includegraphics[scale =0.7]{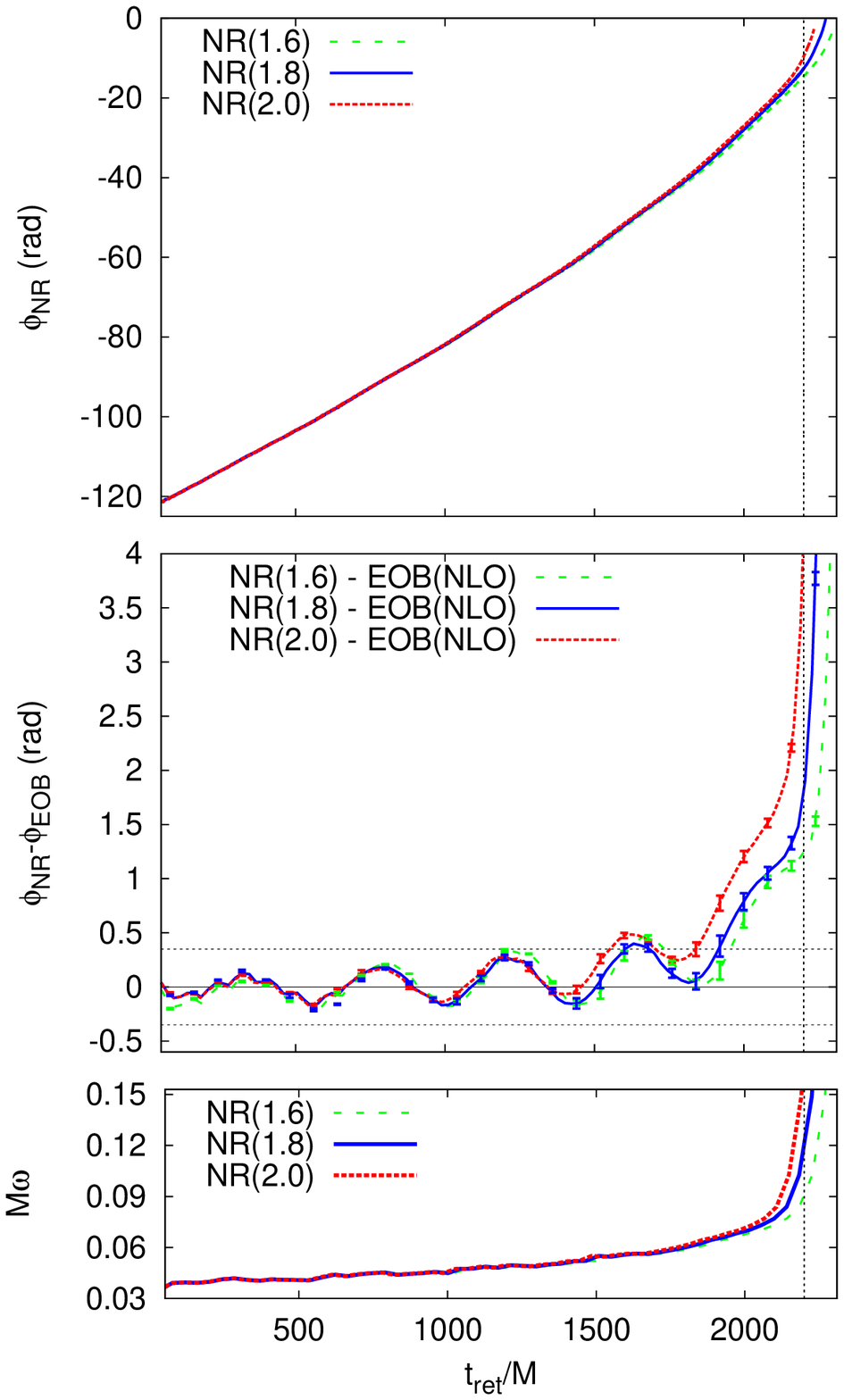}}\\
 \end{tabular}
 \caption{Gravitational-wave phase (top panels), difference between
   $\phi_{\rm{NR}}$ and $\phi_{\rm{EOB}}$ (middle panels), and
   gravitational-wave angular velocity (bottom panels) for APR4 (left
   panels) and H4 (right panels).  EOB(NLO) denotes the effective-one-body
   formalism including the tidal effects up to the next-to leading order.
   We show the curves for the three
   orders of the convergence, $n=(1.6,1.8,2.0)$.
   The horizontal dashed lines in the middle panels denote the
   uncertainty due to the modulation of the numerical data.
   The vertical dashed lines denote the time at contact of the two stars for the
   case of the order of the convergence is $1.8$.}
 \label{evolv-w}
\end{figure*}

\begin{figure}[!ht]
\hspace{-1.0cm} \rotatebox{0}{\includegraphics[width=80mm,clip]{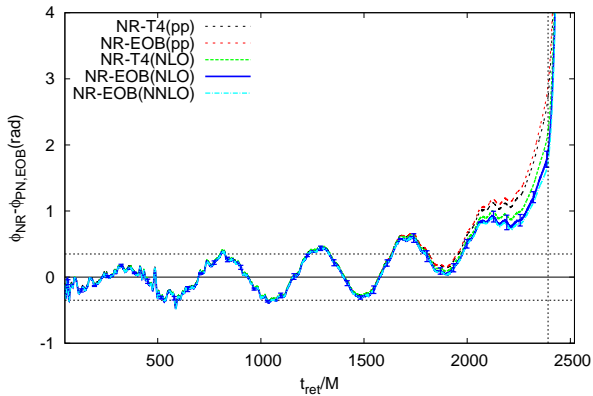}}
 \caption{ Phase difference between $\phi_{\rm{NR}}$ and$\phi_{\rm{PN/EOB}}$
           for APR4. In this figure, we set the order of the
           convergence to be 1.8 for obtaining the extrapolated
           gravitational-wave phase. T4(pp) and EOB(pp) denote
           the Taylor T4 approximant and the effective-one-body
           formalism without the tidal corrections.
           Here, we only display the error bars of the curve
           NR-EOB(NLO). However, other curves also have the
           same amount of the error.}
 \label{ana_eq}
\end{figure}

\begin{figure}[!h]
 \hspace{-1.0cm}\rotatebox{0}{\includegraphics[width=80mm,clip]{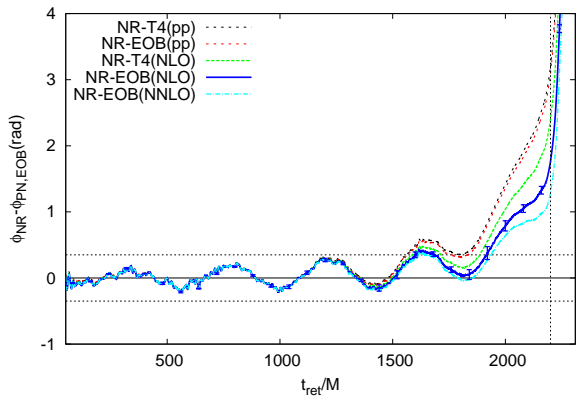}}
  \caption{The same as Fig.~\ref{ana_eq} but for H4. }
 \label{ana_eq1}
\end{figure}

\begin{figure}[!h]
 \hspace{-1.0cm}\rotatebox{0}{\includegraphics[width=80mm,clip]{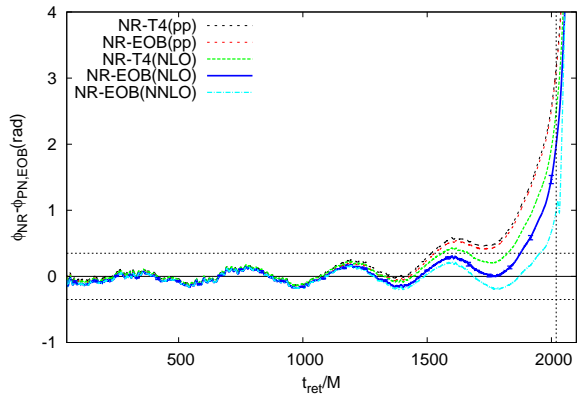}}
  \caption{The same as Fig.~\ref{ana_eq} but for MS1. }
 \label{ana_eq2}
\end{figure}

We compare $\phi_{\rm{NR}}$ with $\phi_{\rm{PN}}$ and
$\phi_{\rm{EOB}}$ which are calculated in several levels of the
approximation.  The middle panels of Fig.~\ref{evolv-w} show the
difference between $\phi_{\rm{NR}}$ and $\phi_{\rm{EOB}}$ in which
tidal effects up to the next-to-leading order are included.  The error
bars show the dispersion associated with the least-square fitting in
the extrapolation procedure, which is defined by Eq.~(41).  The three
curves show the gravitational-wave phase difference for the three
assumed orders of the convergence $n=(1.6, 1.8, 2.0)$.
The phase differences are modulated with an amplitude about $0.4$~radian
due to the unphysical orbital eccentricity.
We regard this modulation as a systematic error, 
which is denoted by the horizontal dashed lines.
For comparison, we also plot the gravitational-wave angular velocity
evolution in the bottom of Fig.~\ref{evolv-w}.
Figure~\ref{ana_eq}--\ref{ana_eq2}
shows the difference between $\phi_{\rm{NR}}$ and $\phi_{\rm{PN/EOB}}$
which are calculated in several levels of the approximation. 

In the early stage of the inspiral, i.e., before the time
$t_{\rm{ret}}/M\lesssim 1500~(M\omega \lesssim 0.054)$ for APR4,
$t_{\rm{ret}}/M\lesssim 1300~(M\omega \lesssim 0.05)$ for H4, and
$t_{\rm{ret}}/M\lesssim 1200~(M\omega \lesssim 0.049)$ for MS1,
we find that
$\phi_{\rm{NR}}$ is consistent with $\phi_{\rm{PN}}$ and
$\phi_{\rm{EOB}}$ for any levels of the approximation.  In this stage,
the point-particle approximation works well and the PN and EOB
approaches appear to describe NS-NS inspirals well even if tidal
effects are not taken into account. It is difficult to verify a clear
signature of the tidal effects in this early stage due to the
modulation of the numerical data and the weakness of the tidal
effects. 

After the early stage of the inspiral, the binary system proceeds to a
tidally-dominated inspiral phase where the tidal interaction becomes
strong.  In this late inspiral stage, one can see that the difference
between $\phi_{\rm{NR}}$ and $\phi_{\rm{PN/EOB}}$ gradually
increases. $\phi_{\rm{NR}}$ is larger than
$\phi_{\rm{PN/EOB}}$ for a given moment, implying
binaries in NR simulations
evolve faster than those in the PN and EOB calculations.

After the time $t_{\rm{ret}}/M\sim 2300~(M\omega\sim0.09)$ for APR4,
$t_{\rm{ret}}/M \sim2100~(M\omega \sim 0.08$) for H4, and
$t_{\rm{ret}}/M \sim1950~(M\omega \sim 0.075$) for MS1,
the difference
between $\phi_{\rm{NR}}$ and $\phi_{\rm{EOB}}$ rapidly increases with
increasing time.  This corresponds to the transition from the inspiral
to the plunge.  The orbital angular velocity when two neutron stars
come into contact, $M\omega_{\rm{contact}}$, is approximately defined
by~\cite{damour12}
\begin{eqnarray}
M\omega_{\rm{contact}} = 2\left(\frac{M_{A}}{M}\frac{1}{C_{A}}+
\frac{M_{B}}{M}\frac{1}{C_{B}}\right)^{-3/2}.
\end{eqnarray}
At the moment of this contact, the accumulated gravitational-wave
phase difference between $\phi_{\rm{NR}}$ and $\phi_{\rm{EOB}}$
including the tidal effects up to the next-to-leading order is $\sim
3.3$~radian ($2.5\%$) for APR4, $\sim 1.9$~radian ($1.7\%$) for H4,
and $\sim 2.1$~radian ($2.0\%$) for MS1.  If the correction is up to
the next-to-next-to-leading order, the phase difference is $\sim
2.6$~radian ($2.0\%$) for APR4, $\sim 1.4$~radian ($1.3\%$) for H4,
and $\sim 1.1$~radian ($1.1\%$) for MS1. 
We find that the EOB approach including the tidal effects up to the
next-to-next-to-leading order yields currently the best model for the late stage of NS-NS inspirals.
However, the tidal effects are still underestimated for the final inspiral orbit even for
the best model of the analytic approaches.
Thus, for constructing better waveforms in NS-NS inspirals, this rapid evolution of the
gravitational-wave phase should be taken into account.

We compare our results of MS1, of which the compactness is 
$C=0.138$, with the results of Ref.~\cite{bernuzzi11},
for which a neutron-star with $C=0.14$ is employed.
Both results show that the difference between the extrapolated
gravitational-wave phase of NR and that of T4 without tidal effects
is about 3 radian at contact~($M\omega \sim 0.1$),
in the case that the alignment is performed after the first orbit~\cite{bernuzzi11}.
Therefore our results are consistent with the results of Ref.~\cite{bernuzzi11}.

Figure~\ref{snap-APR-mw0042} and Fig.~\ref{snap-H4-mw0042} show the snapshots of the density contour
of NS-NS inspirals for APR4 and H4. Here we focus on Fig.~\ref{snap-H4-mw0042}
as an example. The upper left panel of
Fig.~\ref{snap-H4-mw0042} plots the density profile of the binary at
an early inspiral stage.  At this time, the neutron stars have a
spherical shape.  The ellipticity of the neutron star, which is
defined by the ratio of the semi-major axis $a_{1}$ to the semi-minor
axis $a_{2}$ of the star on the equatorial plane, is approximately
unity.

The upper right panel of Fig.~\ref{snap-H4-mw0042}, at $M\omega =
0.056$, plots the configuration of the binary at which the tidal
effects seem to be small but cannot be neglected. For this plot, the
ellipticity of the star is $\sim 1.05$.  After this time, the binary
system proceeds to the tidally-dominated inspiral phase.  In the lower
left panel of Fig.~\ref{snap-H4-mw0042}, two neutron stars are
obviously deformed due to the strong tidal fields; the ellipticity of
the neutron star is $\sim 1.17$.  The snapshot around the plunge is
shown in the lower right panel of Fig.~\ref{snap-H4-mw0042}.  Soon
after the onset of the plunge, the two neutron stars contact and the
ellipticity is $\sim 1.23$.  
In addition, one can see the appearance
of the dynamical tidal lag. It appears even in the absence of viscous dissipation,
because the shape of the star cannot follow the rapid change of the
tidal potential (see, e.g., Ref.~\cite{lai94_2}).
Therefore, the adiabatic approximation for the tidally induced quadrepole
moment determined by Eq.~(1) breaks down after the appearance of the 
dynamical tidal lag. 
In such a case, the tidal deformability should be evaluated with the formalism beyond the adiabatic approximation, 
which is formulated in Ref.~\cite{ferrari12}.

We also note that the ellipticity of the neutron star
rapidly increases after the stage at $M\omega=0.056$.  This evolution
of the ellipticity is consistent with the semianalytic, Newtonian
results~\cite{lai94}.  Note that the value of the ellipticity
defined here depends
on the coordinate system.  We here assumed that the coordinate
distortion is small because of our choice of the
spatial gauge condition.

\begin{figure*}[!h]
 \begin{tabular}{l l}
 \rotatebox{0}{\includegraphics[scale =0.7]{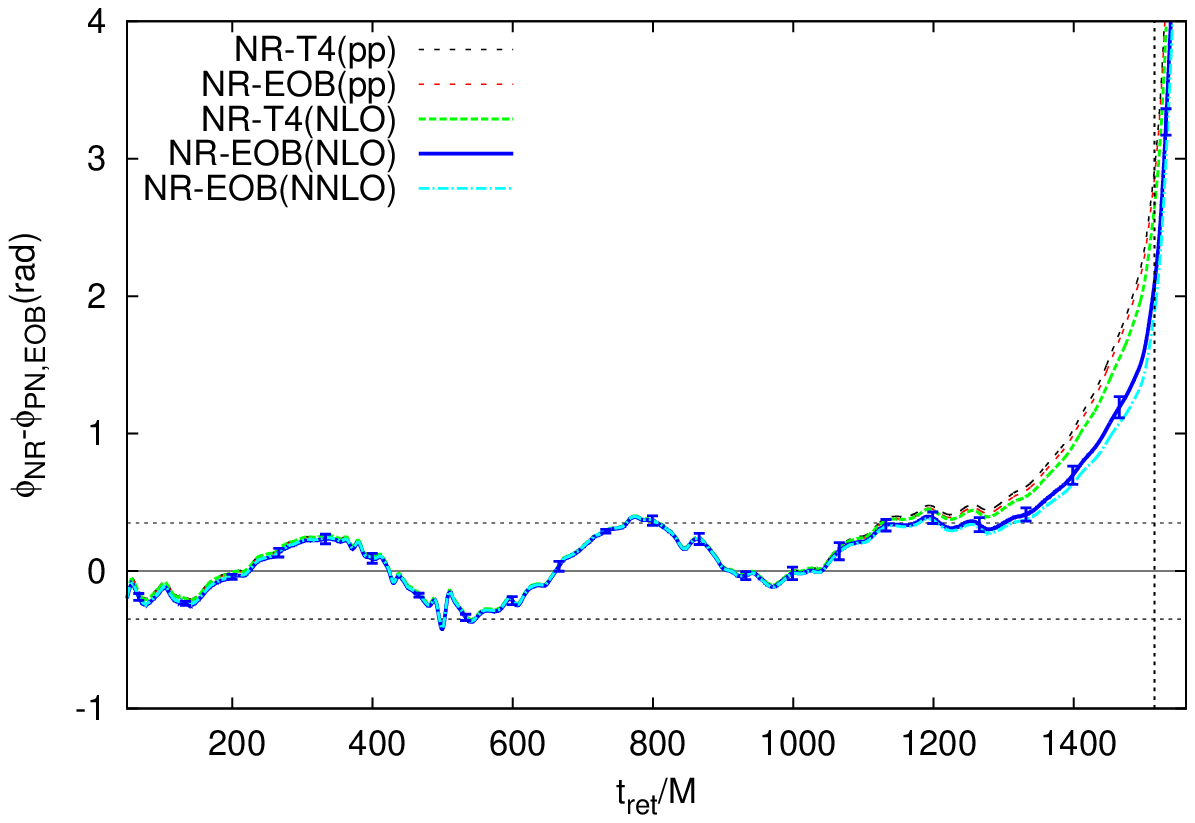}}
 \rotatebox{0}{\includegraphics[scale =0.7]{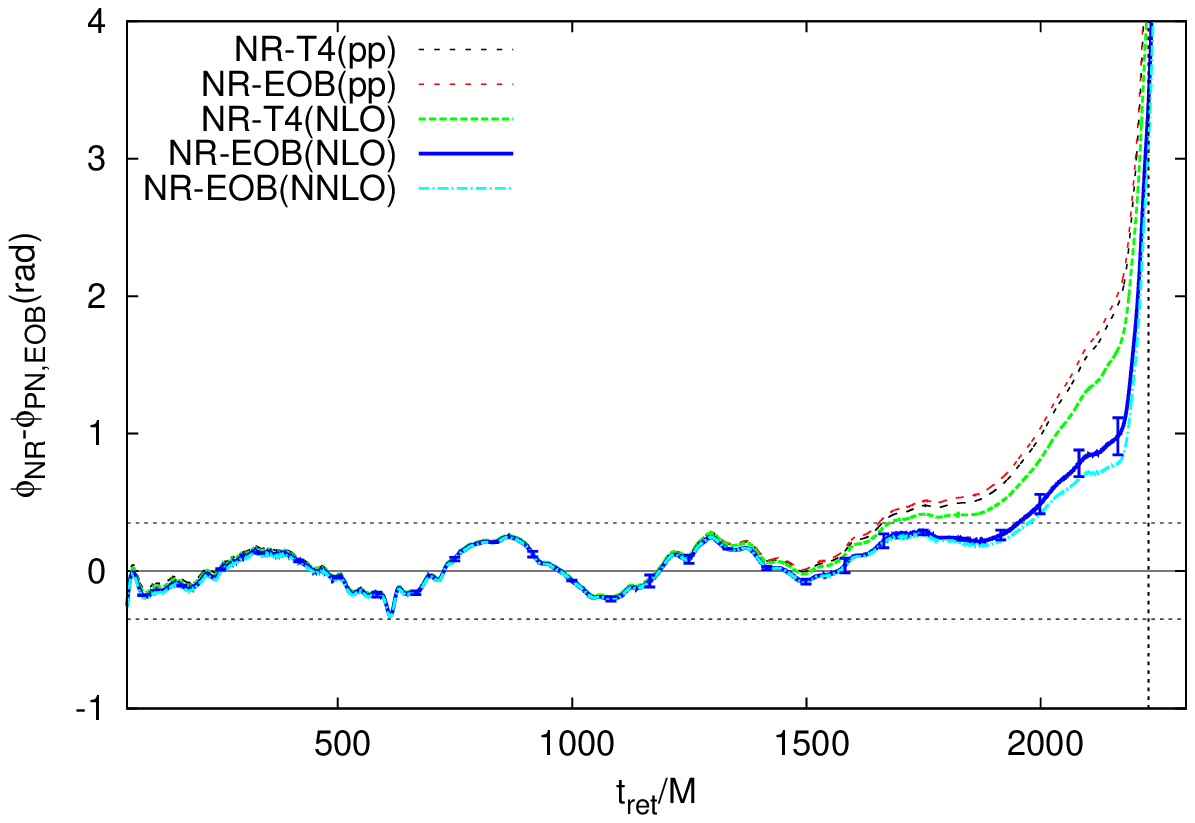}}
 \end{tabular}
 \caption{Difference between $\phi_{\rm{NR}}$ and $\phi_{\rm{PN/EOB}}$
   for APR4-1215 (left panel) and for H4-1215 (right panel). 
   Here we choose $n=1.8$.}
 \label{ana_uneq}
\end{figure*}

Finally, we show the difference between the extrapolated
gravitational-wave phase and the gravitational-wave phase derived in
the PN and EOB approaches for unequal-mass systems APR4-1215 and
H4-1215: see Fig.~\ref{ana_uneq} .  Here, we choose the matching
region $(t_{1},t_{2})=(520M,1020M)$ for APR4-1215 and
$(t_{1},t_{2})=(600M,1080M)$ for H4-1215.  This figure shows that the
feature of the curves for the unequal-mass system is qualitatively the
same as those of the equal-mass system, and the magnitude of the 
phase difference between $\phi_{\rm{NR}}$ and $\phi_{\rm{PN/EOB}}$ 
is also as large as that for the equal-mass case. 

\section{Conclusion}

We have explored the property of gravitational waves emitted in the
late stage of NS-NS inspirals.  To derive the physical waveform from
numerical data obtained by NR simulations, we carefully performed the
radius and resolution extrapolation of the waveforms.  Then, we found
the resolution extrapolation is crucial in our present study.
Specifically, the accumulated difference between the
gravitational-wave phase in the highest grid resolution run and the
resolution-extrapolated gravitational-wave phase is $\sim 8$~radian
for APR4 and $\sim 3$~radian for H4. These values imply that
the simulation for the NS-NS inspiral with more compact neutron stars has
worse convergence. For the simulation with more compact neutron stars
, one needs to perform a higher resolution simulation to derive
an accurate waveform.
Therefore a sophisticated procedure for the extrapolation is needed
to derive an accurate waveform for compact neutron stars.
We found that the time-rescaling is a robust prescription 
for deriving the resolution-extrapolated gravitational-wave
phase $\phi_{\rm{NR}}$.

We have compared $\phi_{\rm{NR}}$ with $\phi_{\rm{PN/EOB}}$, which are
derived from the Taylor T4 approximant of the PN formalism and the EOB
formalism.  Both of the analytic approaches are capable of including
the tidal effects.  We found that $\phi_{\rm{NR}}$ is consistent with
$\phi_{\rm{PN/EOB}}$ in the early part of the inspiral.  On the other
hand, in the very late part of the inspiral, $\phi_{\rm{NR}}$ evolves
more rapidly than $\phi_{\rm{PN/EOB}}$.  The EOB approach including
the tidal corrections up to the next-to-next-to-leading order is
currently the best approach for describing the late stage of NS-NS
inspirals. However, the estimated accumulated difference between
$\phi_{\rm{NR}}$ and $\phi_{\rm{EOB}}$ is $\sim 2.6$~radian for APR4,
$\sim 1.9$~radian for H4, and $\sim 1.1$~radian for MS1 at the moment
of contact of the two neutron stars.  We conclude that the tidal
effects are still underestimated in the EOB approach including the
tidal corrections up to the next-to-next-to-leading order.  
We also found that this result is independent of the EOS and mass ratio.

Here we make a comparison of our results with the earlier results.
We find the absence of the large amplification of the tidal
effects in the early inspiral phase suggested in Ref.~\cite{baiotti11},
if the resolution extrapolation is taken into account.
In the late inspiral phase, the amplification of the tidal
effects is observed. This agrees with the earlier 
results~\cite{baiotti11,bernuzzi11}. In particular, for the MS1~($C\sim 0.14$),
the value of the phase difference between the extrapolated gravitational-wave
phase and that of Taylor T4 without tidal effects is consistent with
the results of Ref.~\cite{bernuzzi11}, which employed the neutron
star with $C=0.14$.

\begin{figure*}[!h]
 \begin{center}
   \begin{tabular}{l l}
\hspace{-1.0cm}\rotatebox{0}{\includegraphics[scale=0.55]{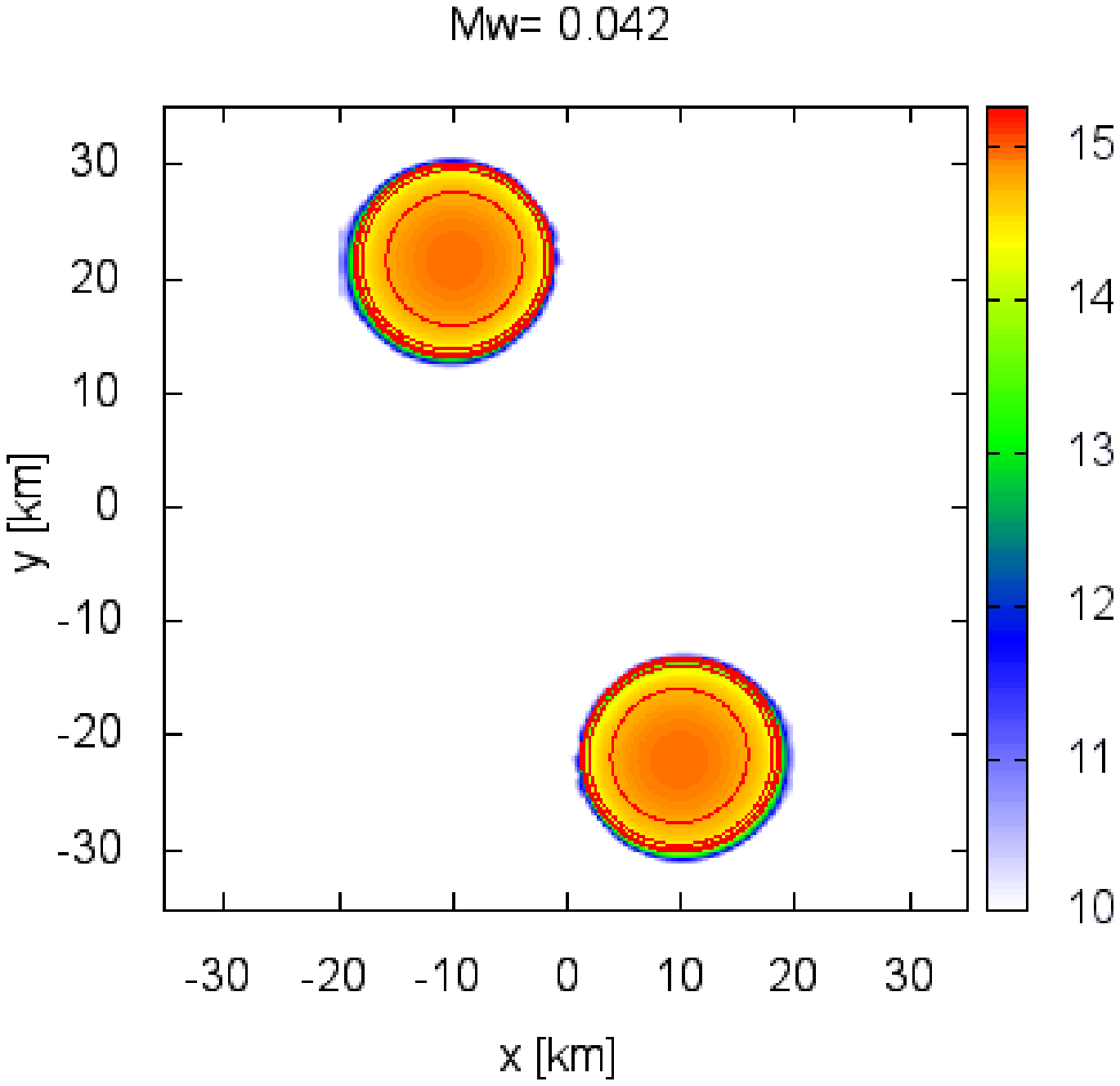}}
\hspace{1.5cm}\rotatebox{0}{\includegraphics[scale=0.55]{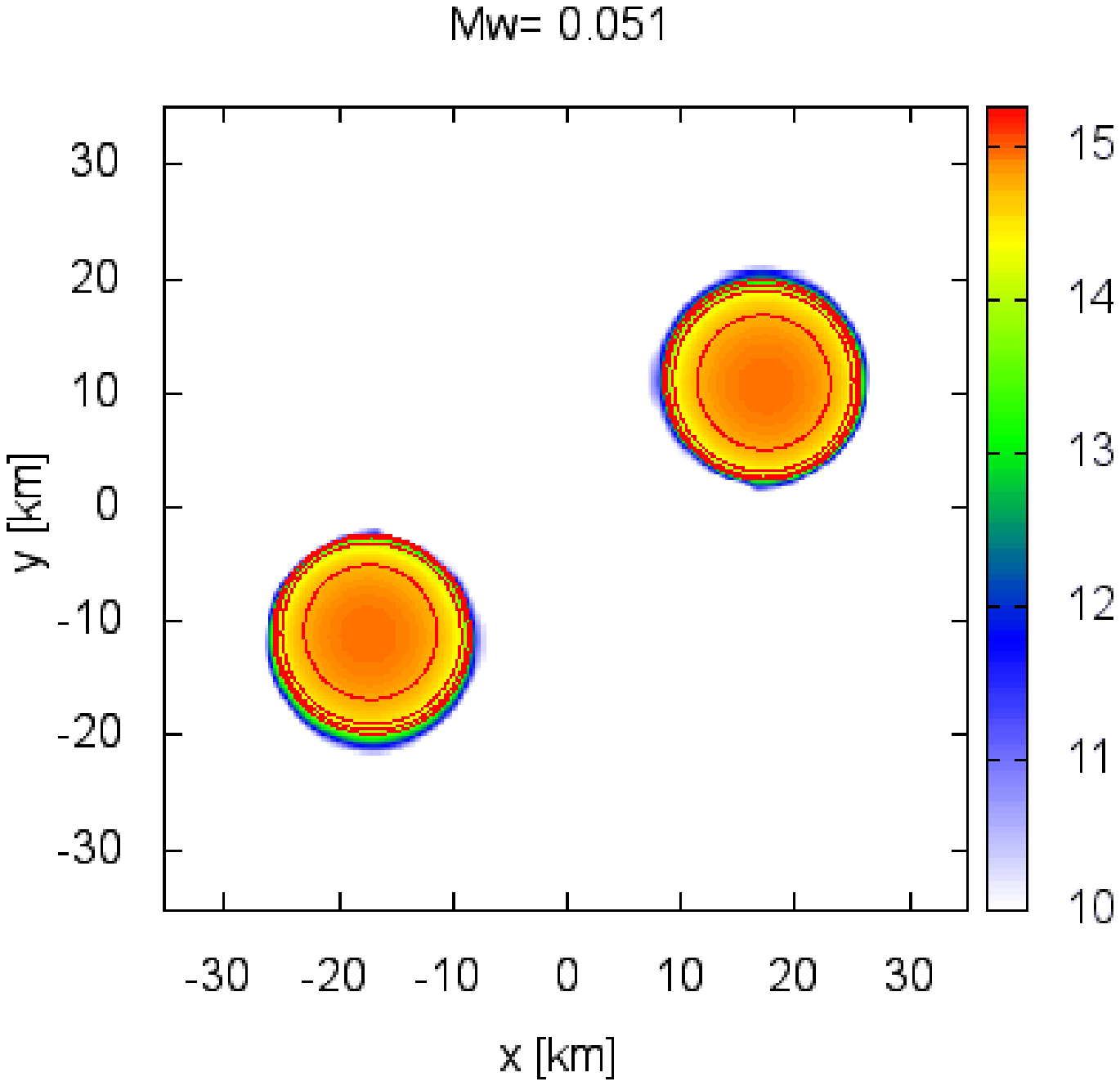}}\\
\hspace{-1.0cm}\rotatebox{0}{\includegraphics[scale=0.55]{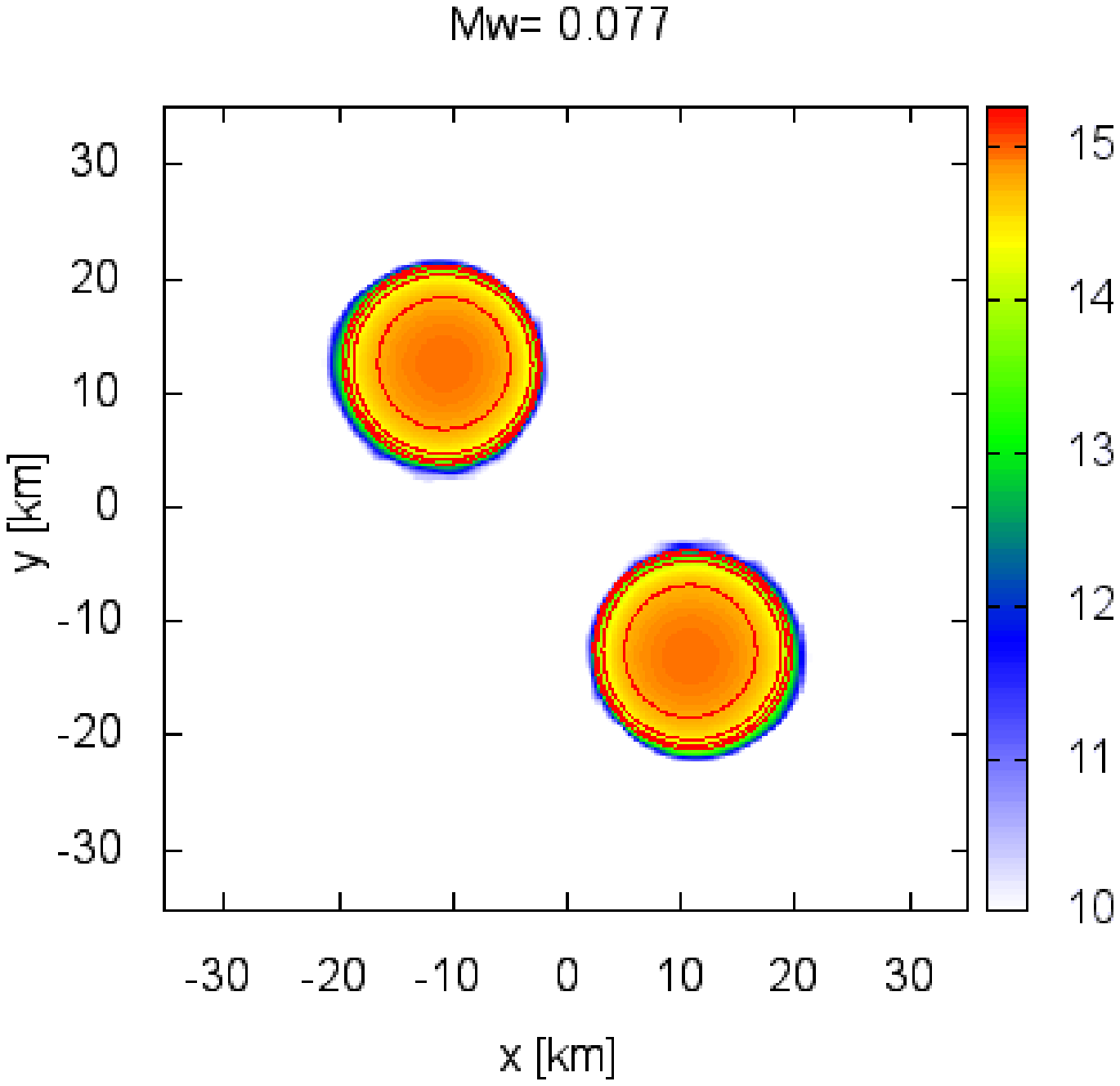}}
\hspace{1.5cm}\rotatebox{0}{\includegraphics[scale=0.55]{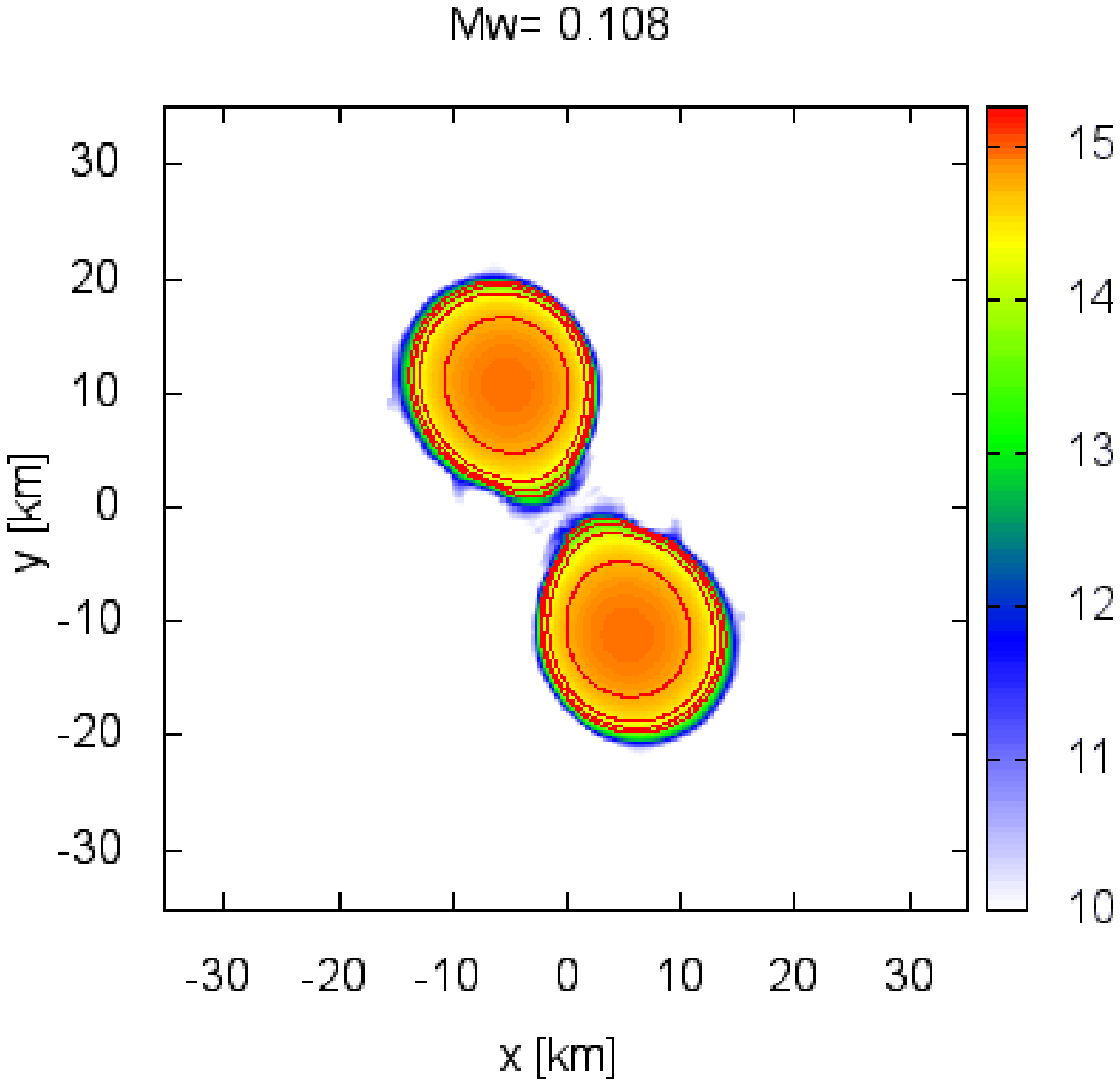}}
   \end{tabular}
 \end{center}
 \caption{Snapshots of the orbital-plane density profile of NS-NS
   binaries in close orbits for APR4. The color denotes the density
   in units of $\rm{log}\rho (\rm{g/cm^{3}})$.}
 \label{snap-APR-mw0042}
\end{figure*}

\begin{figure*}[!h]
 \begin{center}
   \begin{tabular}{l l}
\hspace{-1.0cm}\rotatebox{0}{\includegraphics[scale=0.55]{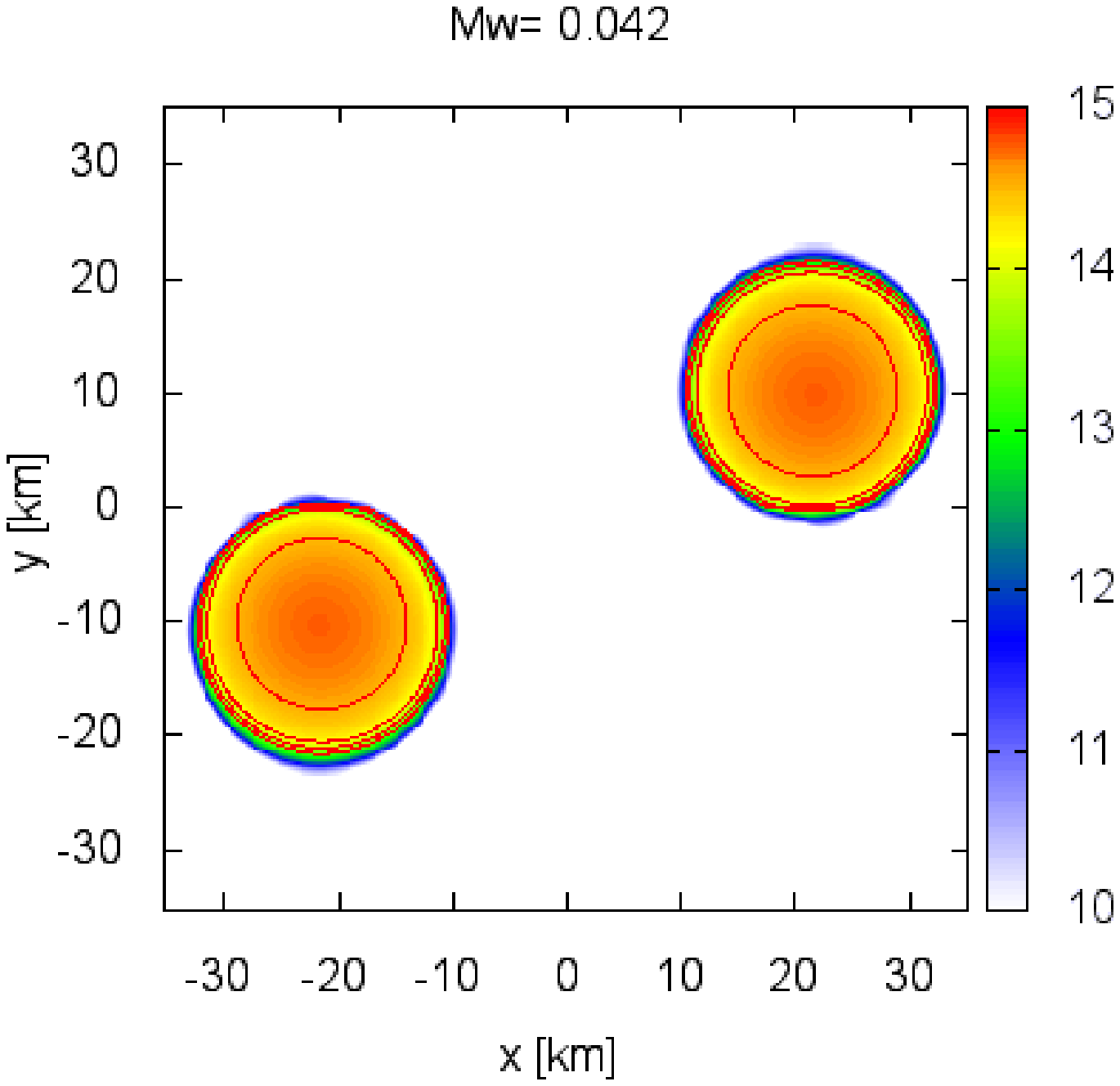}}
\hspace{1.5cm}\rotatebox{0}{\includegraphics[scale=0.55]{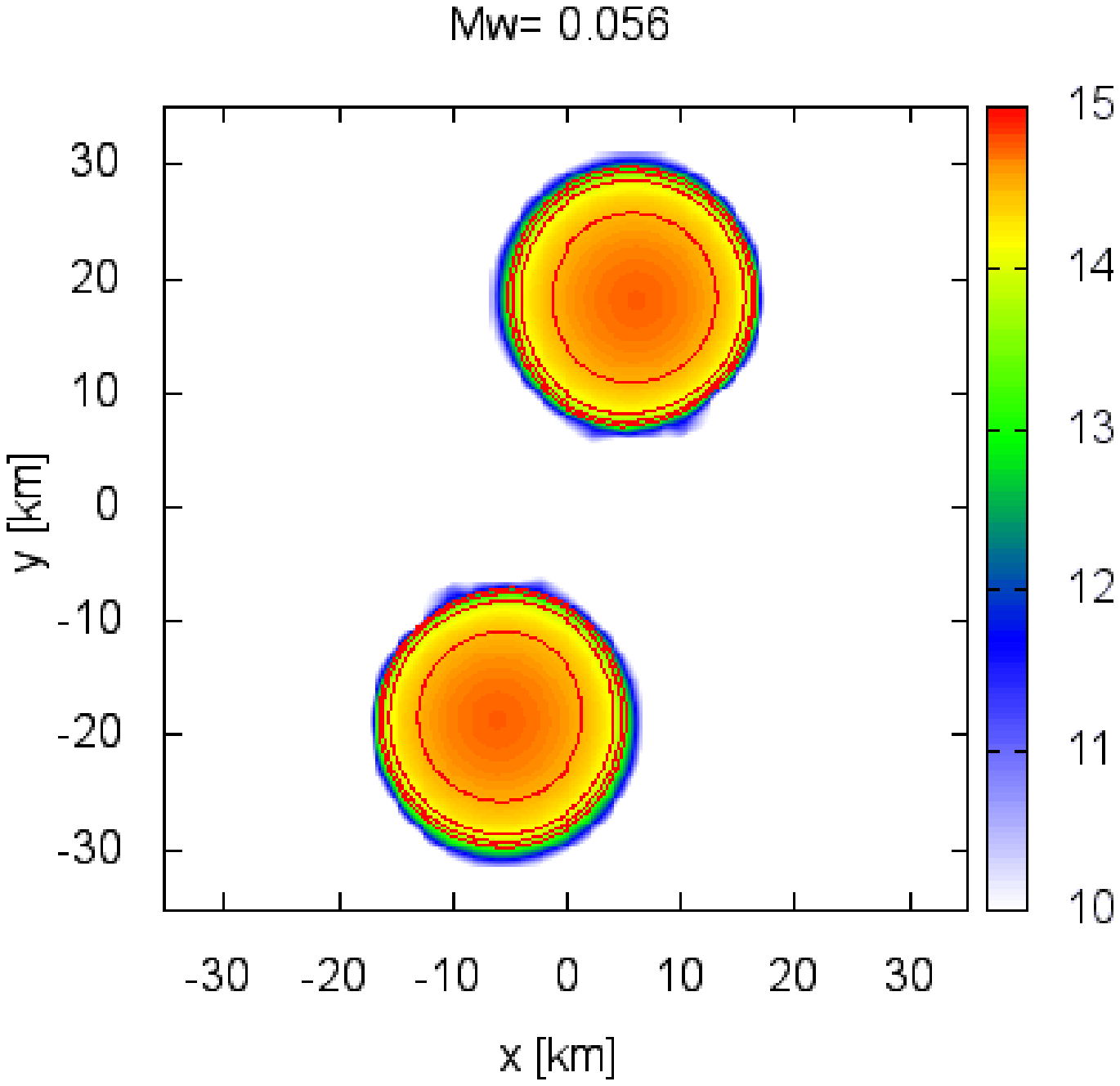}}\\
\hspace{-1.0cm}\rotatebox{0}{\includegraphics[scale=0.55]{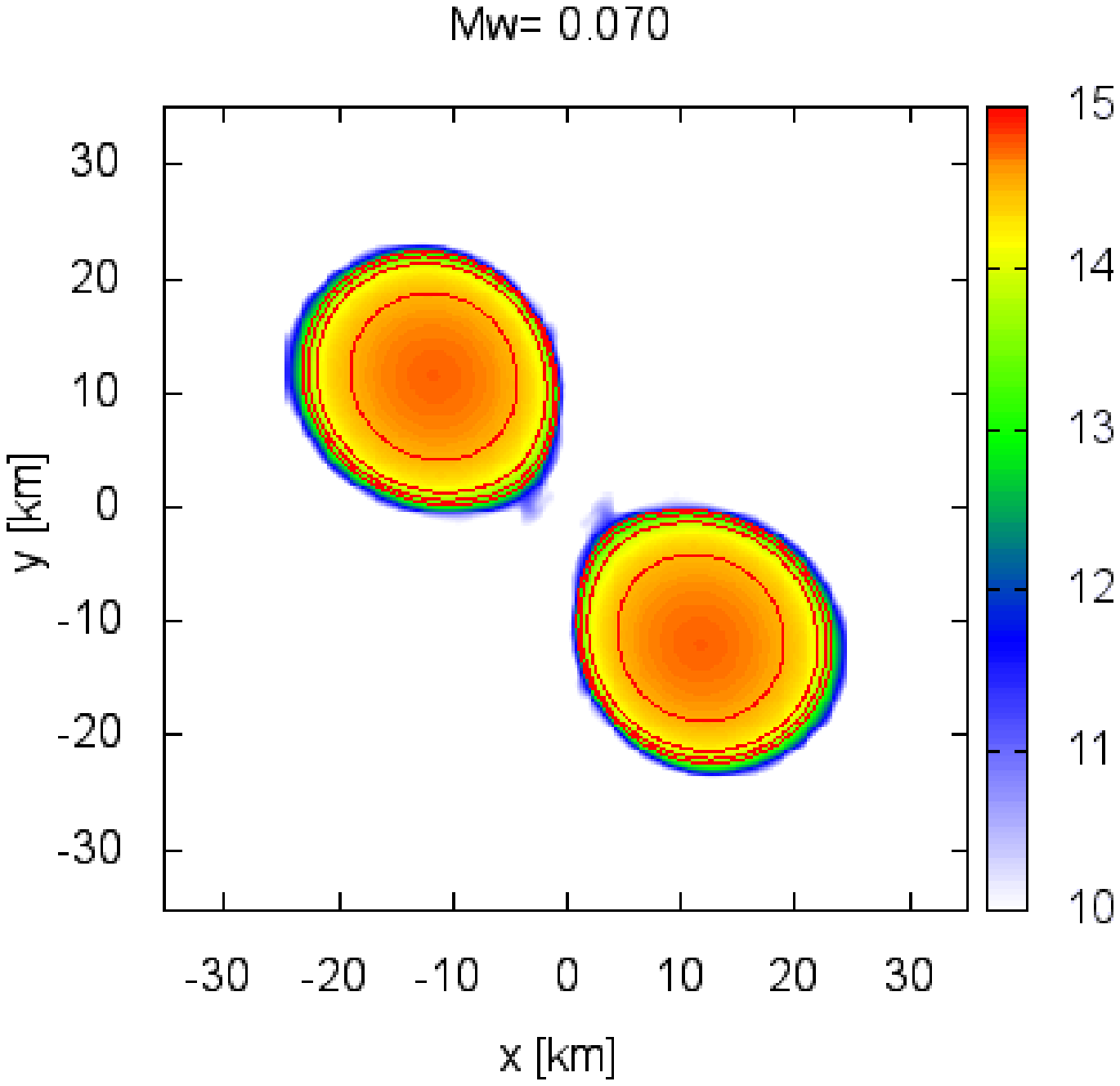}}
\hspace{1.5cm}\rotatebox{0}{\includegraphics[scale=0.55]{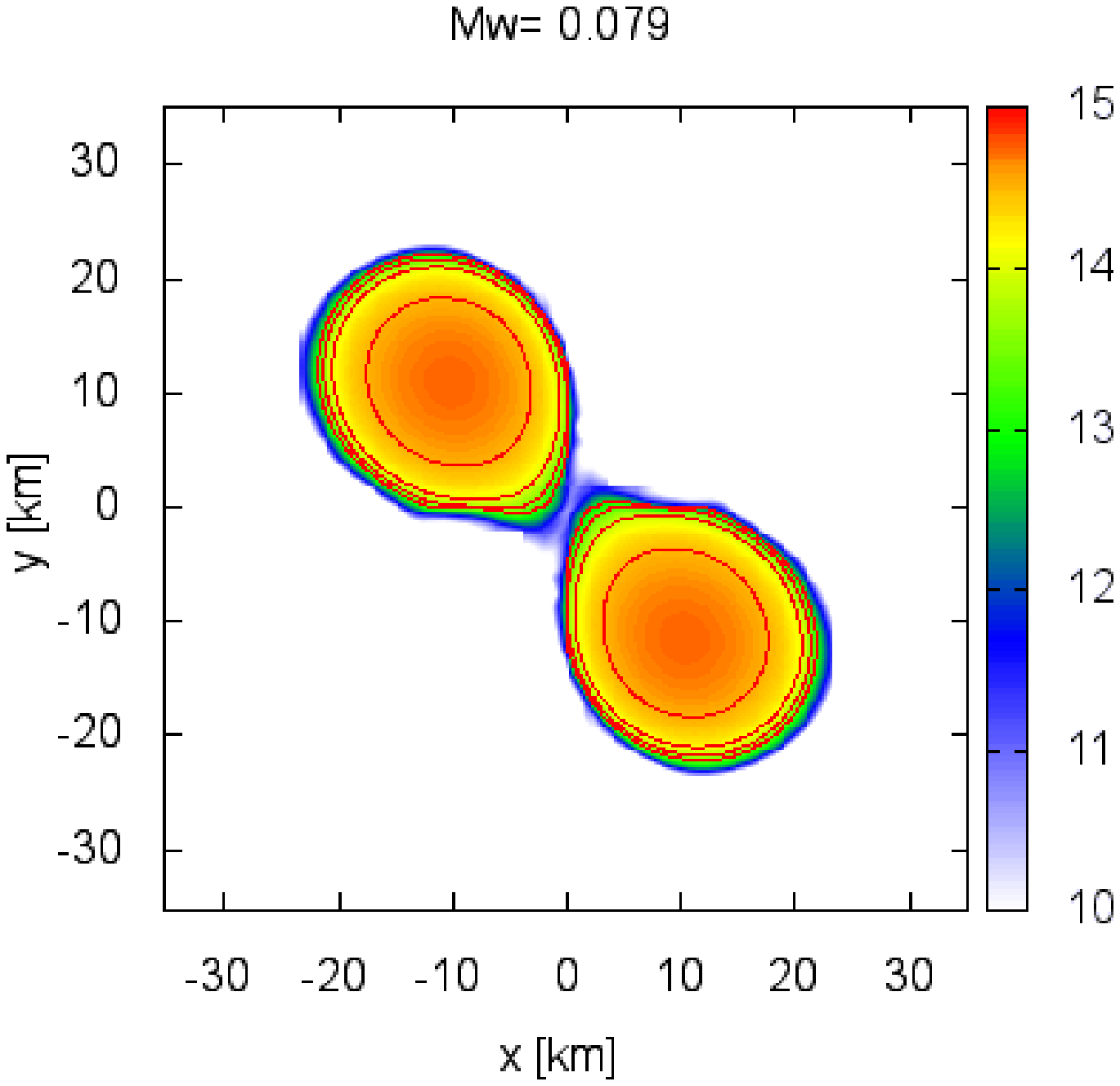}}
   \end{tabular}
 \end{center}
 \caption{The same as Fig.\ref{snap-APR-mw0042} but for H4.}
 \label{snap-H4-mw0042}
\end{figure*}

For extracting the tidal deformability of a neutron star efficiently and
faithfully from a signal of gravitational waves, one has to prepare a
theoretical template of gravitational waves which should be accurate
enough up to the onset of the merger. Our present study suggests that
the EOB approach including up to the next-to-next-to-leading order
tidal correction yields currently the best result. However, for the
final orbit, there is still room for the improvement. In the current
prescription for incorporating the tidal effects, the adiabatic
approximation is assumed for the tidal deformation. For the very close
orbits, however, this approximation breaks down; for example, the
presence of the dynamical tidal lag (which is seen in
Figs.~\ref{snap-APR-mw0042} and \ref{snap-H4-mw0042}) cannot be
reproduced. If a more sophisticated formalism in which such effects

\noindent
are taken in account can be developed, the accuracy of the analytic
\noindent
modeling could be improved (see e.g., Ref~\cite{ferrari12}).

There is also the issue on the side of numerical relativity.  For
constructing an analytic model of the gravitational waveform in 
NS-NS inspirals with a high accuracy, one always calibrates the model waveform by comparing
it with the NR result. This implies that a high-accuracy numerical
waveform is necessary. To achieve a high accuracy, one needs to
perform more accurate simulations than the present ones. One of the
keys for improving the accuracy is to reduce the orbital eccentricity
in the future.

\begin{acknowledgments}
We thank A. Buonanno, K. Kiuchi, J. S. Read, and Y. Sekiguchi for
useful discussions and comments, and B. D. Lackey and G. Faye for
suggesting post-Newtonian and effective one body formalisms for our
analysis.  This work was supported by Grant-in-Aid for Scientific
Research (21340051, 24740163, 21684014), by Grant-in-Aid for Scientific Research
on Innovative Area (20105004), by the Grant-in-Aid of JSPS, by HPCI
Strategic Program of Japanese MEXT.  The work of Hotokezaka is
supported by the Grant-in-Aid of JSPS.
\end{acknowledgments}

\end{document}